\documentclass[Journal]{IEEEtran}

\ifCLASSINFOpdf \usepackage[pdftex]{graphicx} \else \fi

\usepackage[cmex10]{amsmath}
\usepackage{amssymb}
\usepackage{floatrow}
\usepackage{verbatim}
\usepackage{mathdots}
\usepackage{mathtools}

\usepackage{psfrag}

\usepackage{cite}
\usepackage{hyperref}
\usepackage[caption=false,font=footnotesize]{subfig}
\usepackage{stfloats}
\fnbelowfloat
\usepackage{mathrsfs}
\usepackage[utf8]{inputenc}
\usepackage[T1]{fontenc}

\usepackage{enumitem}
\usepackage{circuitikz}
\usepackage{dsfont}
\usepackage{tabularx}

\newcounter{muni}
\newenvironment{remunerate}
               {\begin{list}{{\upshape 
               \arabic{muni}.}}{\usecounter{muni}
                \setlength{\leftmargin}{0pt}
                \setlength{\itemindent}{5pt}}}{\end{list}}

\makeatletter
\newcommand{\labitem}[2]{%
\def\@itemlabel{#1}
\item
\def\@currentlabel{#1}\label{#2}}
\makeatother

\begin{document}

\ctikzset{bipoles/resistor/height=.23}
\ctikzset{bipoles/resistor/width/.initial=.6}
\ctikzset{bipoles/capacitor/height/.initial=.5}
\ctikzset{bipoles/capacitor/width/.initial=.13}
\ctikzset{bipoles/length=1.65cm}
\ctikzset{bipoles/cuteinductor/coils=4}
\ctikzset{bipoles/cuteinductor/lower coil height=.15}
\ctikzset{bipoles/cuteinductor/width=.65}
\ctikzset{bipoles/cuteinductor/height=.25}
\ctikzset{bipoles/cuteinductor/coil aspect=0.25}

\title{Minimal series-parallel network realizations of bicubic impedances}
\author{Timothy H. Hughes \thanks{Timothy H.\ Hughes is with the Department of Mathematics University of Exeter, Penryn Campus, Penryn, Cornwall, TR10 9EZ, UK, e-mail: t.h.hughes@exeter.ac.uk.}}


\newtheorem{theorem}{Theorem}
\newtheorem{lemma}[theorem]{Lemma}
\newtheorem{proposition}[theorem]{Proposition}
\newtheorem{corollary}[theorem]{Corollary}
\newtheorem{definition}[theorem]{Definition}
\newtheorem{remark}[theorem]{Remark}
\newtheorem{problem}[theorem]{Problem}
\providecommand{\abs}[1]{\left\lvert#1\right\rvert}

\newtheorem{thmapp}{Theorem}[section]
\newtheorem{lemapp}[thmapp]{Lemma}
\newtheorem{defapp}[thmapp]{Definition}

\maketitle

\begin{abstract}                          
An important open problem in the synthesis of passive controllers is to obtain a passive network that realizes an arbitrary given impedance function and contains the least possible number of elements. This problem has its origins in electric circuit theory, and is directly applicable to the cost-effective design of mechanical systems containing the inerter. Despite a rich history, the problem can only be considered solved for networks that contain at most two energy storage elements, and in a small number of other special cases. In this paper, we solve the minimal network realization problem for the class of impedances realized by series-parallel networks containing at most three energy storage elements. To accomplish this, we develop a novel continuity-based approach to eliminate redundant elements from a network.
\end{abstract}


\section{Introduction}

In this paper, we provide minimal network realizations for the class of impedance functions realized by series-parallel networks containing three energy storage elements and a finite number of resistive elements. Using the force-current analogy, the results in the paper are directly applicable to the design of both electric circuits without transformers, and mechanical controllers containing the inerter \cite{mcs02}. From a practical perspective, the mechanical applications are of particular interest, as space and cost constraints motivate the design of mechanical systems that have a simple configuration (e.g., series-parallel) and contain the least possible number of elements. Indeed, low complexity networks of the type considered in the paper are relevant to vibration control in a diverse range of application areas, including automotive vehicles \cite{chen_14}; railway vehicles \cite{fucheng_10, fucheng_12, jiang_vsd_12}; buildings \cite{fucheng_07, LNW_SVS, ZJN_LVSMS}; motorcycle steering systems \cite{Limebeer_steering2, Limebeer_Steering}; and aircraft landing systems \cite{YJN_MLGSS}. The use of passive components is particularly beneficial when there are safety or regulatory requirements, or when access to a reliable energy source cannot be guaranteed.

This paper is inspired by the modern control-based framework for the design of mechanical systems pioneered in \cite{mcs02}, which is reminiscent of the behavioral notion of control by interconnection \cite{JWBAOIS}. In this framework, the design problem is to synthesize a dynamical controller taken from a broadly defined class (in this case, the class of series-parallel networks), to interconnect with a given environment (e.g., the connection of a vibration absorber between consecutive floors of a building). Familiar system-theoretic notions such as realizability and minimality take on a different perspective in this framework, resulting in a number of interesting open questions \cite{kalman2010, HugJSmOp, camwb, JW_LTIRLC}. Indeed, despite the relative simplicity of series-parallel networks, it is not known how to obtain a minimal series-parallel realization for an arbitrary impedance function.

To date, much of the literature on this topic has focussed on impedance functions of McMillan degree two (biquadratics). The paper \cite{LIN} provided minimal series-parallel realizations for the class of biquadratic impedances realized by series-parallel networks containing two energy storage elements and a finite number of resistors. An algebraic characterisation of this class of impedances was subsequently provided in \cite{JiangSmith11}. Somewhat surprisingly, it was shown in \cite{HugSmSP} that six energy storage elements are required in order to realize a so-called \emph{biquadratic minimum function} with a series-parallel network. Analogous results to those in \cite{LIN, JiangSmith11, HugSmSP} have also been obtained for general resistor-inductor-capacitor (RLC) networks (i.e., without the series-parallel restriction) in \cite{Reichert69, JZJ_TR, HugGMF}. We also note the papers \cite{ChenSmith09, Chen13, Chen15}, which consider the class of impedances realized by mechanical (or RLC) networks containing one inerter (capacitor), one damper (resistor) and a finite number of springs (inductors).

In contrast, there have been relatively few systematic studies of impedance functions of McMillan degree three (bicubics). A notable exception is the recent paper \cite{ZJ_ERBI}, which considered the realization of a special class of bicubic functions termed \emph{essential regular}, and provided a minimal series-parallel realization for each impedance in this class. However, not all series-parallel networks with three energy storage elements have essential regular impedances. Thus, the purpose of the present paper is to solve the minimal realization problem for the entire class of impedance functions realized by series-parallel networks containing three energy storage elements and a finite number of resistors. In particular, we provide an explicit minimal network realization for every single impedance within this class. In contrast to existing studies, we introduce a novel continuity-based technique to eliminate resistors from non-minimal networks in order to obtain the aforementioned minimal series-parallel network realizations. Specifically, we first obtain non-minimal network realizations for the entire class. Then, for each such network, we show that the element values within the network can be varied  continuously without changing the impedance of the network in order to replace one of the resistors with either a short or open circuit (see the proof of Theorem \ref{thm:bcn1013}). 

For comparison with existing literature, we state our results in terms of electrical series-parallel networks. However, as described earlier, the results have direct practical relevance to the design of mechanical controllers using the force-current analogy. The motivation for this paper in the context of passive mechanical control is further elaborated on in Section \ref{sec:pmc}, where the main results are illustrated by a practical example from \cite{fucheng_9b}. The rest of the paper is then structured as follows. In Section \ref{sec:ncm}, we present a formal approach to network classification that serves to simplify the statement and proof of our main results. Sections \ref{sec:ac} and \ref{sec:blbqi} summarize relevant established results in the passive network synthesis literature. Our new results then follow in Sections \ref{sec:nmgs} and \ref{sec:mgsbc}. In Theorems \ref{thm:z12nmgs}--\ref{thm:z21nmgs} and Lemmas \ref{lem:z30mgs}--\ref{lem:z03mgs}, we provide network realizations for the class of impedances realized by series-parallel networks containing at most three energy storage elements. The realizations are non-minimal (in the number of elements used) except in cases when all energy storage elements are of the same type. The main contributions of the paper then follow from Section \ref{sec:mgsbc} onwards. In Lemmas \ref{lem:z30mgs}--\ref{lem:z03mgs} and Theorems \ref{thm:z12mgs}--\ref{thm:z21mgs}, we prove that any impedance that can be realized by a series-parallel network containing at most three energy storage elements (with no constraint on the number of resistors) can also be realized by a series-parallel network containing at most three energy storage elements and at most four resistors. Moreover, Theorems \ref{thm:mr3re}--\ref{thm:bqn14} provide a routine and computationally tractable method for obtaining an explicit minimal series-parallel network realization for any given impedance from this class.

Finally, our notation is as follows. The real numbers are denoted $\mathbb{R}$; $\mathbb{R}^{n}$ denotes $n$-dimensional Euclidean space; $\Re{(z)}$ denotes the real part of the complex number $z$; and $\mathbb{R}[s]$ (resp., $\mathbb{R}(s)$, $\mathbb{R}[u,v]$) denotes the univariate polynomials (resp., univariate rational functions, bivariate polynomials) with coefficients from $\mathbb{R}$. We make extensive use of the correspondence between polynomials and polynomial functions, where we denote the argument with a tilde whenever the latter interpretation is being applied. In other words: (i) any given $f \in \mathbb{R}[s]$ also defines a continuous function from $\mathbb{R}$ to $\mathbb{R}$ whose value for any given $\tilde{x} \in \mathbb{R}$ corresponds to the evaluation of the polynomial at $\tilde{x}$, denoted $f(\tilde{x})$; and (ii) if $f, g \in \mathbb{R}[s]$, then $s$ is to be interpreted as an indeterminate in the equation $f(s) = g(s)$, implying that $f$ and $g$ are identical polynomials or equivalently that $f(\tilde{x}) = g(\tilde{x})$ for all $\tilde{x} \in \mathbb{R}$. A similar convention is followed for multivariate polynomials.

\section{Series-parallel networks: classification and minimality}
\label{sec:ncm}

We begin this section with some technical preliminaries on network classification that simplify the statement and proof of our results. This formalism is similar to \cite{JiangSmith11, HugSmSP, MS_LCGA}. We then formally state the problem considered in this paper.

We define a series-parallel network in the manner of \cite{Riordan_42}. Specifically, an individual resistor, inductor or capacitor is a series-parallel network, and a network is series-parallel if it is either a series or parallel connection of two series-parallel networks. The impedance $Z$ of a series (resp., parallel) connection of two networks $N_{1}$ and $N_{2}$ with impedances $Z_{1}$ and $Z_{2}$ satisfies $Z = Z_{1} + Z_{2}$ (resp., $1/Z = 1/Z_{1} + 1/Z_{2}$). For a given series-parallel network $N$ with impedance $Z(s)$, there exists a series-parallel network whose impedance is $Z(1/s)$ (denoted $N^{i}$), and a series-parallel network whose impedance is $1/Z(s)$ (denoted $N^{d}$). The network $N^{i}$ is obtained by replacing inductors (with impedance $Xs$) with capacitors (with impedance $X/s$) and vice-versa; and $N^{d}$ is obtained by interchanging series and parallel connections and inverting the impedances of each element (which again replaces inductors with capacitors and vice-versa). In particular, $(N^{i})^{d} = (N^{d})^{i}$, and we denote this network by $N^{p}$. We will also define network classes (denoted $\mathcal{N}_{1}, \mathcal{N}_{2}$, etc) as sets that contain all networks of a given fixed structure, in addition to networks obtained by replacing certain resistors in this structure with open or short circuits (see, e.g., Fig.\ \ref{fig:bq1c1l}). Finally, for a given network class $\mathcal{N}$, we let $\mathcal{N}^{i} = \lbrace N \mid \exists N_{b} \in \mathcal{N} \text{ with } N = N_{b}^{i}\rbrace$, and the network classes $\mathcal{N}^{d}$ and $\mathcal{N}^{p}$ are defined similarly.

To formally state our results, and compare them with the existing literature, we next introduce the concepts of generic network classes and minimal generating sets.

The impedance of a given RLC network always takes the form of a ratio of two polynomials
\begin{align}
p(s) &= p_{n}s^{n} + p_{n-1}s^{n-1} + \cdots + p_{1}s + p_{0}, \text{ and} \nonumber\\
q(s) &= q_{n}s^{n} + q_{n-1}s^{n-1} + \cdots + q_{1}s + q_{0}, \label{eq:pqd}
\end{align}
for some integer $n$ whose value does not exceed the number of energy storage elements in the network. Here, no generality is lost in requiring either $p_n \neq 0$ or $q_n \neq 0$, and the coefficients $p_{0}, \ldots , p_{n}, q_{0}, \ldots , q_{n}$ are all polynomial functions in the network's element values (inductances, capacitances, etc.) that can be obtained from Kirchhoff's tree formula \cite{HMS_GEN}. Following \cite{HMS_GEN}, we call the set of impedances realized by a given network class $\mathcal{N}$ the \emph{realizability set} of $\mathcal{N}$, which can be characterised by the vector of coefficients $(p_{0}, \ldots , p_{n}, q_{0}, \ldots , q_{n})$ and viewed as a (semi-algebraic) subset of $\mathbb{R}^{2n+2}$. For any given $N \in \mathcal{N}$, the dimension\footnote{We define the dimension of a semi-algebraic set $\mathcal{S}$ in accordance with \cite{ARAG} as the largest $d$ such that there exists a one-to-one smooth map from the open cube $(-1,1)^d \subset \mathbb{R}^d$ into $\mathcal{S}$.} of the realizability set of $\mathcal{N}$ is at most one greater than the number of elements in $N$ \cite[Lemma 2]{HMS_GEN}, and $\mathcal{N}$ is called generic if every single network in $\mathcal{N}$ contains (strictly) fewer elements than the dimension of the realizability set of $\mathcal{N}$ \cite[Definition 1]{HMS_GEN}. It follows from \cite[Lemma 2]{HMS_GEN} that almost all networks from a given generic network class are \emph{minimal} in the sense that their impedance cannot be realized by a network containing strictly fewer elements.\footnote{More precisely, the set of impedances in the realizability set that can be realized with strictly fewer elements is a subset of the realizability set whose codimension is at least one.}

\begin{definition}[Generating/ minimal generating sets]
\label{defn:gs}
Let $\mathcal{Z}_{m,n}$ be the set of impedances realized by series-parallel networks containing at most $m$ capacitors and $n$ inductors; and let $\mathcal{Z}_{M} = \cup_{m,n \mid m + n = M} \mathcal{Z}_{m,n}$ be the set of impedances realized by series-parallel networks containing at most $M$ energy storage elements. We call $\mathcal{G}$ a \emph{generating set} for $\mathcal{Z}_{m,n}$ (resp., $\mathcal{Z}_{M}$) if (i) $\mathcal{G}$ is a set of series-parallel networks, each containing at most $m+n$ (resp., $M$) energy storage elements; and (ii) for every single $Z \in \mathcal{Z}_{m,n}$ (resp., $Z \in \mathcal{Z}_{M}$), there exists a network $N \in \mathcal{G}$ whose impedance is $Z$. We call $\mathcal{G}$ a \emph{minimal generating set} for $\mathcal{Z}_{m,n}$ (resp., $\mathcal{Z}_M$) if (i) $\mathcal{G}$ is a generating set; and (ii) $\mathcal{G}$ is the union of generic network classes.
\end{definition}

\begin{problem}
\label{prob:fgs}
Given $\mathcal{Z}_{m,n}$ (resp., $\mathcal{Z}_{M}$) as in Definition \ref{defn:gs}:
\begin{enumerate}[label=(\alph*)]
\item Find a generating set for $\mathcal{Z}_{m,n}$ (resp., $\mathcal{Z}_{M}$).\label{nl:fgsa}
\item Find a minimal generating set for $\mathcal{Z}_{m,n}$ (resp., $\mathcal{Z}_{M}$).\label{nl:fgsb}
\end{enumerate}
\end{problem}

Clearly, any solution to problem \ref{prob:fgs}\!\ref{nl:fgsb} also solves  \ref{prob:fgs}\!\ref{nl:fgsa}. Solutions to problem \ref{prob:fgs}\!\ref{nl:fgsb} for the sets $\mathcal{Z}_{0,n}$ and $\mathcal{Z}_{m,0}$ are provided by the so-called \emph{Cauer form} for all integer $m$ and $n$, for which explicit realizations are provided in \cite{HugCF}. Moreover, problem \ref{prob:fgs}\!\ref{nl:fgsb} has also been solved for the case $M = 2$. As discussed in \cite{HugJSmOp}, both the increased complexity of algebraic manipulations, and the growth in the number of candidate network structures, present considerable barriers to extending these results to cases with $M > 2$. Thus, in contrast with existing approaches, we develop (in the proof of Theorem \ref{thm:bcn1013}) a novel continuity-based argument to solve problem \ref{prob:fgs}\!\ref{nl:fgsb} in the case $M = 3$. A natural extension of the present paper is to consider cases with greater numbers of energy storage elements. Given the algebraic complexity of the problem, it is envisaged that similar continuity based arguments will be essential for solving these cases.

\section{Algebraic criteria for network realizations}
\label{sec:ac}

The impedance of a given RLC network always takes the form $Z = p/q$ for some polynomials $p, q$ as in (\ref{eq:pqd}). Here, no generality is lost by assuming that at least one of $p_{n}, q_{n}$ is non-zero, and $p, q$ are coprime (equivalently, the McMillan degree of $p/q$ is $n$). In \cite{HugSmAI}, several necessary algebraic conditions were presented for a function $Z = p/q$ to be the impedance of an RLC network. These relate to the Sylvester matrices and their determinants:\footnote{We will also consider the case in which the coefficients are themselves polynomials, i.e., $p_n, p_{n-1}, \ldots , q_n, q_{n-1}, \ldots \in \mathbb{R}[u]$, in which case $F_k(p,q) \in \mathbb{R}[u]$ for $k = 0, 1, \ldots , n-1$. Then, for any given $\tilde{x} \in \mathbb{R}$, $F_k(p,q)(\tilde{x})$ denotes the evaluation of this polynomial at $\tilde{x}$.}
\def\matriximg{%
  \begin{matrix}
    q_{n}& q_{n-1}& q_{n-2}& \cdots\\
p_{n}& p_{n-1}& p_{n-2}& \cdots\\
0& q_{n}& q_{n-1}& \cdots\\
0& p_{n}& p_{n-1}& \cdots\\
0& 0& q_{n}& \cdots\\
\vdots& \vdots& \vdots& \ddots
   \end{matrix}
}%
\begin{align*} \mathcal{S}_{i}(p,q) &:=
  \left. \left[\vphantom{\matriximg} \kern-\nulldelimiterspace \right. \overbrace{\matriximg}^{\text{$i$ columns}} \right.
  \left. \left.\vphantom{\matriximg}\right] \kern-2\nulldelimiterspace \right\} \text{\scriptsize $i$ rows}, (i = 1, 2, \ldots ), \\
  F_{k}(p,q) &:= \abs{\mathcal{S}_{2(n-k)}(p,q)}, \hspace{0.15cm} (k = 0, 1, \ldots, n-1).
\end{align*}
Here, $F_0(p,q)$ is proportional to the resultant of $p$ and $q$ (and is equal in magnitude if $p_{n} \neq 0$ and $q_{n} \neq 0$), and $p$ and $q$ have at least $r$ roots in common (counting according to multiplicity) if and only if $F_0(p, q) = \cdots = F_{r-1}(p, q) = 0$ \cite{HugSmAI}. The following result is then shown in \cite[Theorem 8]{HugSmAI}:
\begin{lemma}
\label{lem:ai}
Let $p, q$ in (\ref{eq:pqd}) be coprime, and let $Z = p/q$ be the impedance of an RLC network $N$ containing at most $n$ energy storage elements. Then $F_0(p,q) \neq 0$, and the number of capacitors (resp., inductors) in $N$ is equal to the number of permanences (resp., variations) in sign in the sequence $\lbrace 1, F_{n-1}(p,q), \ldots , F_{1}(p,q), F_0(p,q)\rbrace$.

In any subsequence of zero values, $F_{k}(p,q) \neq 0$, $F_{k-1}(p,q) = F_{k - 2}(p,q) = \ldots = 0$, signs are assigned to the zero values as follows: $\text{sign}(F_{k - j}) = (-1)^{j(j-1)/2}\text{sign}(F_{k})$.
\end{lemma}
\begin{remark}
Let $m_{1}, m_{2}, n_{1}, n_{2}$ be integers with $M = m_{1} + n_{1} = m_{2} + n_{2}$ and $m_{1} \neq m_{2}$. From Lemma \ref{lem:ai}, if $Z \in \mathcal{Z}_{m_{1}, n_{1}}$ has McMillan degree $M$, then $Z \not\in \mathcal{Z}_{m_{2}, n_{2}}$.
\end{remark}

Lemma \ref{lem:ai} can also be stated in terms of the \emph{Bezoutian matrix} associated with the polynomials $p$ and $q$:
\begin{definition}
\label{def:bezd}
Let $p, q$ be as in (\ref{eq:pqd}). Then $\mathcal{B}(q,p)$ is the matrix whose entries $\mathcal{B}_{ij}$ satisfy\footnote{We will also consider the case in which the coefficients in $p$ are themselves polynomials, i.e., $p_n, p_{n-1} \ldots \in \mathbb{R}[u]$, in which case $\mathcal{B}(q,p)$ is a matrix of polynomials in the indeterminate $u$. Then, for any given $\tilde{x} \in \mathbb{R}$, $\mathcal{B}(q,p)(\tilde{x})$ denotes the evaluation of this polynomial matrix at $\tilde{x}$.\label{fn:bmp}}
\begin{equation*}
\frac{q(z)p(w) - p(z)q(w)}{z - w} = \sum_{i=1}^{n}{\sum_{j=1}^{n}{\mathcal{B}_{ij}z^{i-1}w^{j-1}}}.
\end{equation*}
\end{definition}
From \cite[Section 6]{HugSmAI}, $F_{k}(p,q)$ is equal to the determinant formed from the final $n{-}k$ rows and columns of $\mathcal{B}(q,p)$. In particular, we note the following:
\begin{corollary}
\label{cor:dbei}
Let $p,q$ in (\ref{eq:pqd}) be coprime, and let $Z = p/q$ be the impedance of an RLC network $N$ containing at most $n$ energy storage elements. Then $\abs{\mathcal{B}(q,p)} = F_0(p,q) \neq 0$, and $N$ contains an even number of inductors if and only if $\abs{\mathcal{B}(q,p)} = F_0(p,q) > 0$.
\end{corollary}

\section{Bilinear and biquadratic impedances}
\label{sec:blbqi}

In this section, we summarize several known results on those impedances that are realized by series-parallel networks containing at most two energy storage elements. These results provide minimal generating sets for $\mathcal{Z}_{0}, \mathcal{Z}_{1}$ and $\mathcal{Z}_{2}$. We also prove a couple of lemmas on the properties of the sets $\mathcal{Z}_{1}$ and $\mathcal{Z}_{2}$. These lemmas will be used in Section \ref{sec:mgsbc} to obtain a minimal generating set for $\mathcal{Z}_{3}$.

A classical result in passive network synthesis is that the impedance of any network that contains only one type of energy storage element can always be realized by the so-called Cauer canonical networks. These networks provide minimal generating sets for $\mathcal{Z}_{0,k}$ and $\mathcal{Z}_{k,0}$ for $k = 0, 1, 2, \ldots$.\footnote{To see that the Cauer canonical network classes are generic, it suffices to show that there is a one to one function from the impedance parameters to the element values and scaling parameters (denoted $E_{1}, \ldots , E_{m}, c$ in \cite[equation (3)]{HMS_GEN}). This is easily shown from results in \cite{HugCF}.} In the case of $\mathcal{Z}_{0,0}$, a minimal generating set is given by the set of all resistors (see Fig.\ \ref{fig:c0r}). The cases $\mathcal{Z}_{0,1}$ and $\mathcal{Z}_{1,0}$ are covered in the following lemma:
\begin{lemma}
\label{lem:blgs}
Consider the network classes in Figs.\ \ref{fig:c0r}--\ref{fig:bl1inags}. The following hold.
\begin{enumerate}
\item $\mathcal{N}_{1} \cup \mathcal{N}_{2}$ is a minimal generating set for $\mathcal{Z}_{1,0}$.
\item $\mathcal{N}_{1} \cup \mathcal{N}_{2a}$ is a minimal generating set for $\mathcal{Z}_{1,0}$.
\item $\mathcal{N}_{1} \cup \mathcal{N}_{3}$ is a minimal generating set for $\mathcal{Z}_{0,1}$.
\item $\mathcal{N}_{1} \cup \mathcal{N}_{3a}$ is a minimal generating set for $\mathcal{Z}_{0,1}$.
\end{enumerate}
In particular, if $Z \in \mathcal{Z}_1$ has McMillan degree zero, then $Z$ is realized by the resistor $N_1$ in Fig.\ \ref{fig:c0r} with $R_1 = Z \geq 0$. If, on the other hand, $Z = f/g$ where
\begin{equation}
f(s) = f_1s + f_0 \text{ and } g(s) = g_1 s + g_0 \label{eq:fgdef}
\end{equation}
are coprime and at least one of $g_0, g_1$ are non-zero, then
\begin{enumerate}
\item if $Z \in \mathcal{Z}_{1,0}$, then $F_0(f,g) > 0$, $f_1, g_0$ and $g_1$ have the same sign\footnote{Here, and throughout, we say a set of real numbers have the same sign if either all are non-negative or all are non-positive \label{fn:ss}}, $g_1 \neq 0$, and $Z$ is the impedance of $N_2$ in Fig.\ \ref{fig:bl1in} when the element values are as indicated in that figure's caption.
\item if $Z \in \mathcal{Z}_{0,1}$, then $F_0(f,g) < 0$, $f_0, g_0$ and $g_1$ have the same sign, $g_0 \neq 0$, and $Z$ is the impedance of $N_3$ in Fig.\ \ref{fig:bl1in} when the element values are as indicated in that figure's caption.
\end{enumerate}
\end{lemma}

\tikzset{ r1/.pic={
\ctikzset{bipoles/thickness=1}
\node[anchor=center] at (1,0) (O) {};
\draw (O) to[short,*-] ++(0.3,0)
to[R=$R_{1}$] ++(1.7,0)
to[short,-*] ++(0.3,0);}}

\begin{figure}[!b]
\centering
\normalsize
\begin{tikzpicture}[scale=0.57, every node/.style={transform shape}, font=\Large]
\pic at (0,0) [] {r1};
\node at (-1.6,-0.0) {$\text{(i) } R_{1} \geq 0$};
\end{tikzpicture}
\caption{Network $N_{1}$. We define $\mathcal{N}_{1}$ as the set of all networks of the form of $N_{1}$ that satisfy condition (i).}
\label{fig:c0r}
\end{figure}
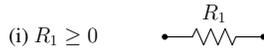

\tikzset{ cbl1/.pic={
\ctikzset{bipoles/thickness=1}
\node[anchor=center] at (1,0) (O) {};
\draw (O) to[short,*-] ++(0.3,0)
to[R=$R_{1}$] ++(1.7,0)
to[short] ++(0,0.7)
to[R=$\frac{1}{G_{2}}$] ++(1.7,0)
to[short] ++(0,-0.7)
to[short,-*] ++(0.3,0)
($(O)+(2.0,0)$) to[short] ++(0,-0.7)
to[C=$\frac{1}{C_{1}s}$] ++(1.7,0)
to[short] ++(0,0.7);}}

\tikzset{ lbl1/.pic={
\ctikzset{bipoles/thickness=1}
\node[anchor=center] at (1,0) (O) {};
\draw (O) to[short,*-] ++(0.3,0)
to[R=$R_{1}$] ++(1.7,0)
to[short] ++(0,0.7)
to[R=$\frac{1}{G_{2}}$] ++(1.7,0)
to[short] ++(0,-0.7)
to[short,-*] ++(0.3,0)
($(O)+(2.0,0)$) to[short] ++(0,-0.7)
to[L=$L_{1}s$] ++(1.7,0)
to[short] ++(0,0.7);}}

\begin{figure}[!b]
\centering
\normalsize
\begin{tikzpicture}[scale=0.57, every node/.style={transform shape}, font=\Large]
\node at (-1.7,-0.2) {$\begin{array}{rl}\text{(i)}& R_{1} \geq 0\\
\text{(ii)}& G_{2} \geq 0\\
\text{(iii)(a)}& C_{1} > 0 \\
\text{(iii)(b)}& L_{1} > 0
\end{array}$};
\node at (1.3,1.2) {$N_{2}$};
\pic at (0,0) [] {cbl1};
\node at (7.0,1.2) {$N_{3}$};
\pic at (5.7,0) [] {lbl1};
\end{tikzpicture}
\caption[]{Networks $N_{2}$ and $N_{3}$. We define $\mathcal{N}_{2}$ (resp., $\mathcal{N}_{3}$) as the set of all networks of the form of $N_{2}$ (resp., $N_{3}$) that satisfy conditions (i), (ii) and (iii)(a) (resp., conditions (i), (ii) and (iii)(b)). 

\quad If the relevant conditions of Lemma \ref{lem:blgs} hold, then $N_2$ and $N_3$ have impedance $f/g$ with $f, g \in \mathbb{R}[s]$ as in (\ref{eq:fgdef}) when the element values are:\\[0.2em]  $\begin{array}{|l|l|}\hline \rule{0pt}{1.1\normalbaselineskip} N_{2} & R_1 = \tfrac{f_1}{g_1}, G_2 = \tfrac{g_0 g_1}{F_0(f,g)}, C_1 = \tfrac{g_1^2}{F_0(f,g)}.\\ \hline \rule{0pt}{1.1\normalbaselineskip} N_{3} & R_1 = \tfrac{f_0}{g_0}, G_2 = \tfrac{-g_0 g_1}{F_0(f,g)}, L_1 = \tfrac{-F_0(f,g)}{g_0^2}.\\ \hline\end{array}$}
\label{fig:bl1in}
\end{figure}
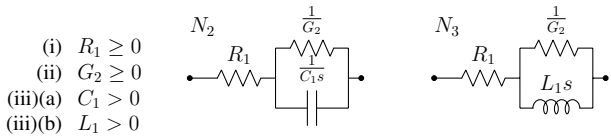

\tikzset{ lbl2/.pic={
\ctikzset{bipoles/thickness=1}
\node[anchor=center] at (1,0) (O) {};
\draw (O) to[short,*-] ++(0.3,0)
to[short] ++(0,0.7)
to[R=$\frac{1}{G_{1}}$] ++(3.4,0)
to[short] ++(0,-0.7)
to[short,-*] ++(0.3,0)
($(O)+(0.3,0)$) to[short] ++(0,-0.7)
to[R=$R_{2}$] ++(1.7,0)
to[L=$L_{1}s$] ++(1.7,0)
to[short] ++(0,0.7);}}

\tikzset{ cbl2/.pic={
\ctikzset{bipoles/thickness=1}
\node[anchor=center] at (1,0) (O) {};
\draw (O) to[short,*-] ++(0.3,0)
to[short] ++(0,0.7)
to[R=$\frac{1}{G_{1}}$] ++(3.4,0)
to[short] ++(0,-0.7)
to[short,-*] ++(0.3,0)
($(O)+(0.3,0)$) to[short] ++(0,-0.7)
to[R=$R_{2}$] ++(1.7,0)
to[C=$\frac{1}{C_{1}s}$] ++(1.7,0)
to[short] ++(0,0.7);}}

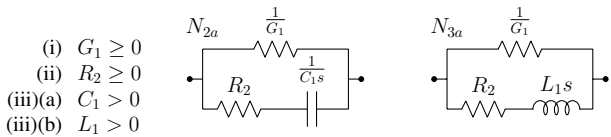
\begin{figure}[!b]
\centering
\normalsize
\begin{tikzpicture}[scale=0.57, every node/.style={transform shape}, font=\Large]
\node at (-1.7,-0.2) {$\begin{array}{rl}\text{(i)}& G_{1} \geq 0\\
\text{(ii)}& R_{2} \geq 0\\
\text{(iii)(a)}& C_{1} > 0 \\
\text{(iii)(b)}& L_{1} > 0
\end{array}$};
\node at (1.3,1.2) {$N_{2a}$};
\node at (7.0,1.2) {$N_{3a}$};
\pic at (0,0) [] {cbl2};
\pic at (5.7,0) [] {lbl2};
\end{tikzpicture}
\caption{Networks $N_{2a}$ and $N_{3a}$. We define $\mathcal{N}_{2a}$ (resp., $\mathcal{N}_{3a}$) as the set of all networks of the form of $N_{2a}$ (resp., $N_{3a}$) that satisfy conditions (i), (ii) and (iii)(a) (resp., conditions (i), (ii) and (iii)(b)).}
\label{fig:bl1inags}
\end{figure}

In fact, any bilinear impedance realized by an RLC (not necessarily series-parallel) network can also be realized by one of the networks in Figs.\ \ref{fig:c0r}--\ref{fig:bl1inags}, in accordance with the following well known result:
\begin{lemma}
\label{lem:blec}
Let the McMillan degree of $Z \in \mathbb{R}(s)$ be less than or equal to one. The following are equivalent:
\begin{enumerate}
\item $Z$ is the impedance of an RLC network.
\item $Z$ is positive-real (i.e., (i) $Z$ is analytic in the open right half plane, and (ii) $\Re{(Z(s_{0}))} \geq 0$ whenever $\Re{(s_{0})} > 0$).\label{nl:blprc}
\item $Z \in \mathcal{Z}_{1}$.
\end{enumerate}
\end{lemma}

A minimal generating set for $\mathcal{Z}_{2}$, which first appeared in \cite{LIN}, is described in the following lemma.
\begin{lemma}
\label{lem:bqmgs}
Let $\mathcal{N}_{1}$--$\mathcal{N}_{9}$ be as in Figs.\ \ref{fig:c0r}--\ref{fig:bq1c1l}. The following hold.
\begin{enumerate}
\item $\mathcal{N}_{1} \cup \mathcal{N}_{2} \cup \mathcal{N}_{4}$ is a minimal generating set for $\mathcal{Z}_{2,0}$.
\item $\mathcal{N}_{1} \cup \mathcal{N}_{3} \cup \mathcal{N}_{5}$ is a minimal generating set for $\mathcal{Z}_{0,2}$.
\item The union of $\mathcal{N}_{1}$--$\mathcal{N}_{3}$ and $\mathcal{N}_{6}$--$\mathcal{N}_{9}$ is a minimal generating set for $\mathcal{Z}_{1,1}$.
\end{enumerate}
\end{lemma}

\begin{IEEEproof}
It is shown in \cite{MS_LC} that the network classes $\mathcal{N}_{1}$--$\mathcal{N}_{9}$ are generic, and the result then follows from Lemma \ref{lem:ai}, \cite{LIN} and \cite[Theorem 1]{JiangSmith11}.
\end{IEEEproof}

\tikzset{ ccbq1/.pic={
\ctikzset{bipoles/thickness=1}
\node[anchor=center] at (1,0) (O) {};
\draw (O) to[short,*-] ++(0.3,0)
to[R=$R_{1}$] ++(1.7,0)
to[short] ++(0,0.7)
to[R=$R_{2}$] ++(1.7,0)
to[short] ++(0,0.7)
to[R=$\frac{1}{G_{3}}$] ++(1.7,0)to[short] ++(0,-1.4)
($(O)+(3.7,0.7)$) to[short] ++(0,-0.7)
to[C=$\frac{1}{C_{1}s}$] ++(1.7,0)
to[short,-*] ++(0.3,0)
($(O)+(2.0,0)$) to[short] ++(0,-1.4)
to[C=$\frac{1}{C_{2}s}$] ++(3.4,0)
to[short] ++(0,1.4);}}

\tikzset{ llbq1/.pic={
\ctikzset{bipoles/thickness=1}
\node[anchor=center] at (1,0) (O) {};
\draw (O) to[short,*-] ++(0.3,0)
to[R=$R_{1}$] ++(1.7,0)
to[short] ++(0,0.7)
to[R=$R_{2}$] ++(1.7,0)
to[short] ++(0,0.7)
to[R=$\frac{1}{G_{3}}$] ++(1.7,0)
to[short] ++(0,-1.4)
($(O)+(3.7,0.7)$) to[short] ++(0,-0.7)
to[L=$L_{1}s$] ++(1.7,0)
to[short,-*] ++(0.3,0)
($(O)+(2.0,0)$) to[short] ++(0,-1.4)
to[L=$L_{2}s$] ++(3.4,0)
to[short] ++(0,1.4);}}

\begin{figure}[!b]
\centering
\normalsize
\begin{tikzpicture}[scale=0.57, every node/.style={transform shape}, font=\Large]
\node at (1.3,1.8) {$N_{4}$};
\pic at (0,0) [] {ccbq1};
\node at (7.8,1.8) {$N_{5}$};
\pic at (6.5,0) [] {llbq1};
\end{tikzpicture}
\caption[]{Networks $N_{4}$ and $N_{5}$. In $N_{4}$, (i) $R_{1} \geq 0$, (ii) $R_{2} > 0$, (iii) $G_{3} \geq 0$, and (iv)(a) $C_{1}, C_{2} > 0$. In $N_{5}$, conditions (i)--(iii) hold, and (iv)(b) $L_{1}, L_{2} > 0$. We define $\mathcal{N}_{4}$ (resp., $\mathcal{N}_{5}$) as the set of all networks of the form of $N_{4}$ (resp., $N_{5}$) that satisfy conditions (i)--(iii) and (iv)(a) (resp., conditions (i)--(iii) and (iv)(b)).

\quad If the relevant conditions of Lemma \ref{lem:bqrrc} hold, and $\gamma_1, \gamma_2, \gamma_3, \lambda_1, \lambda_2, \lambda_3$ and $\lambda_4$ are as defined in that lemma, then $N_4$ and $N_5$ have impedance $c/d$ with $c, d \in \mathbb{R}[s]$ as in (\ref{eq:cdd}) when the element values are:\\[0.2em]
 $\begin{array}{|l|l|}\hline \rule{0pt}{1.1\normalbaselineskip} N_{4} & R_{1} {=} \tfrac{c_{2}}{d_{2}}, R_{2} {=} \tfrac{\gamma_{3}^{2}}{d_{2}\lambda_{3}}, G_{3} {=} \tfrac{d_{0}\lambda_{3}}{F_{0}(c,d)}, C_{1} {=} \tfrac{-\lambda_{3}^{2}}{\gamma_{3}F_{0}(c,d)}, C_{2} {=} \tfrac{-d_{2}^{2}}{\gamma_{3}}.\\ \hline \rule{0pt}{1.1\normalbaselineskip} N_{5} & R_{1} {=} \tfrac{c_{0}}{d_{0}}, R_{2} {=} \tfrac{\gamma_{1}^{2}}{d_{0}\lambda_{1}}, G_{3} {=} \tfrac{d_{2}\lambda_{1}}{F_{0}(c,d)}, L_{1} {=} \tfrac{\gamma_{1}F_{0}(c,d)}{\lambda_{1}^{2}}, L_{2} {=} \tfrac{\gamma_{1}}{d_{0}^{2}}.\\ \hline\end{array}$}
\label{fig:bq2co2i}
\end{figure}
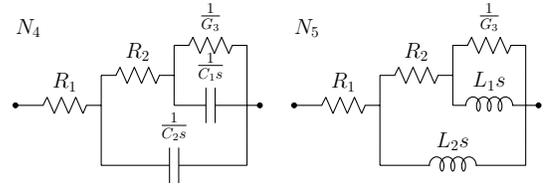

\tikzset{ clbq1/.pic={
\ctikzset{bipoles/thickness=1}
\node[anchor=center] at (1,0) (O) {};
\draw (O) to[short,*-] ++(0.3,0)
to[R=$R_{1}$] ++(1.7,0)
to[short] ++(0,0.7)
to[R=$R_{2}$] ++(1.7,0)
to[short] ++(0,0.7)
to[R=$\frac{1}{G_{3}}$] ++(1.7,0)
to[short] ++(0,-1.4)
($(O)+(3.7,0.7)$) to[short] ++(0,-0.7)
to[C=$\frac{1}{C_{1}s}$] ++(1.7,0)
to[short,-*] ++(0.3,0)
($(O)+(2.0,0)$) to[short] ++(0,-1.4)
to[L=$L_{1}s$] ++(3.4,0)
to[short] ++(0,1.4);}}

\tikzset{ clbq2/.pic={
\ctikzset{bipoles/thickness=1}
\node[anchor=center] at (1,0) (O) {};
\draw (O) to[short,*-] ++(0.3,0)
to[R=$R_{1}$] ++(1.7,0)
to[short] ++(0,0.7)
to[R=$R_{2}$] ++(1.7,0)
to[short] ++(0,0.7)
to[R=$\frac{1}{G_{3}}$] ++(1.7,0)to[short] ++(0,-1.4)
($(O)+(3.7,0.7)$) to[short] ++(0,-0.7)
to[L=$L_{1}s$] ++(1.7,0)
to[short,-*] ++(0.3,0)
($(O)+(2.0,0)$) to[short] ++(0,-1.4)
to[C=$\frac{1}{C_{1}s}$] ++(3.4,0)
to[short] ++(0,1.4);}}

\tikzset{ clbq3/.pic={
\ctikzset{bipoles/thickness=1}
\node[anchor=center] at (1,0) (O) {};
\draw (O) to[short,*-] ++(0.3,0)
to[short] ++(0,-0.7)
to[C=$\frac{1}{C_{1}s}$] ++(1.7,0)
to[short] ++(0,0.7)
to[R=$\frac{1}{G_{2}}$] ++(3.4,0)
to[short,-*] ++(0.3,0)
($(O)+(2.0,-0.7)$) to[short] ++(0,-0.7)
to[R=$R_{3}$] ++(1.7,0)
to[L=$L_{1}s$] ++(1.7,0)
to[short] ++(0,1.4)
($(O)+(0.3,0)$) to[short] ++(0,1.4)
to[R=$\frac{1}{G_{1}}$] ++(5.1,0)
to[short] ++(0,-1.4);}}

\tikzset{ clbq4/.pic={
\ctikzset{bipoles/thickness=1}
\node[anchor=center] at (1,0) (O) {};
\draw (O) to[short,*-] ++(0.3,0)
to[short] ++(0,-0.7)
to[L=$L_{1}s$] ++(1.7,0)
to[short] ++(0,0.7)
to[R=$\frac{1}{G_{2}}$] ++(3.4,0)
to[short,-*] ++(0.3,0)
($(O)+(2.0,-0.7)$) to[short] ++(0,-0.7)
to[R=$R_{3}$] ++(1.7,0)
to[C=$\frac{1}{C_{1}s}$] ++(1.7,0)
to[short] ++(0,1.4)
($(O)+(0.3,0)$) to[short] ++(0,1.4)
to[R=$\frac{1}{G_{1}}$] ++(5.1,0)
to[short] ++(0,-1.4);}}

\begin{figure}[!b]
\centering
\normalsize
\begin{tikzpicture}[scale=0.57, every node/.style={transform shape}, font=\Large]
\node at (1.3,1.8) {$N_{6}$};
\pic at (0,0) [] {clbq1};
\node at (7.8,1.8) {$N_{7}$};
\pic at (6.5,0) [] {clbq2};
\node at (1.3,-2.2) {$N_{8}$};
\pic at (0,-4.0) [] {clbq3};
\node at (7.8,-2.2) {$N_{9}$};
\pic at (6.5,-4.0) [] {clbq4};
\end{tikzpicture}
\caption[]{Networks $N_{6}$--$N_{9}$. In $N_{6}$ and $N_{7}$, (i)(a) $R_{1}, R_{2} \geq 0$, (ii)(a) $G_{3} \geq 0$, and (iii) $C_{1}, L_{1} > 0$. We define $\mathcal{N}_{6}$ (resp., $\mathcal{N}_{7}$) as the set of all networks of the form of $N_{6}$ (resp., $N_{7}$) that satisfy conditions (i)(a), (ii)(a) and (iii). In $N_{8}$ and $N_{9}$, condition (iii) holds, (i)(b) $R_{3} \geq 0$, and (ii)(b) $G_{1}, G_{2} \geq 0$.  We define $\mathcal{N}_{8}$ (resp., $\mathcal{N}_{9}$) as the set of all networks of the form of $N_{8}$ (resp., $N_{9}$) that satisfy conditions (i)(b), (ii)(b) and (iii).

\quad If the relevant conditions of Lemma \ref{lem:bqrrc} hold, and $\gamma_1, \gamma_2, \gamma_3, \lambda_1, \lambda_2, \lambda_3$ and $\lambda_4$ are as defined in that lemma, then $N_6$--$N_9$ have impedance $c/d$ with $c, d \in \mathbb{R}[s]$ as in (\ref{eq:cdd}) when the element values are:\\[0.2em] $\begin{array}{|l|l|}\hline \rule{0pt}{1.1\normalbaselineskip} N_{6} & R_{1} {=} \tfrac{c_{0}}{d_{0}}, R_{2} {=} \tfrac{\gamma_{2}}{d_{0}d_{2}}, G_{3} {=} \tfrac{-d_{2}\lambda_{1}}{F_{0}(c,d)}, C_{1} {=} \tfrac{-d_{2}^{2}\gamma_{1}}{F_{0}(c,d)}, L_{1} {=} \tfrac{\gamma_{1}}{d_{0}^{2}}.\\ \hline \rule{0pt}{1.1\normalbaselineskip} N_{7} &R_{1} {=} \tfrac{c_{2}}{d_{2}}, R_{2} {=} \tfrac{-\gamma_{2}}{d_{0}d_{2}}, G_{3} {=} \tfrac{-d_{0}\lambda_{3}}{F_{0}(c,d)}, C_{1} {=} \tfrac{-d_{2}^{2}}{\gamma_{3}}, L_{1} {=} \tfrac{F_{0}(c,d)}{d_{0}^{2}\gamma_{3}}.\\ \hline \rule{0pt}{1.1\normalbaselineskip} N_{8} & G_{1} {=} \tfrac{d_{0}}{c_{0}}, G_{2} {=} \tfrac{-\gamma_{2}}{c_{0}c_{2}}, R_{3} {=} \tfrac{-c_{2}\lambda_{4}}{F_{0}(c,d)}, C_{1} {=} \tfrac{-\gamma_{1}}{c_{0}^{2}}, L_{1} {=} \tfrac{c_{2}^{2}\gamma_{1}}{F_{0}(c,d)}.\\ \hline \rule{0pt}{1.1\normalbaselineskip} N_{9} & G_{1} {=} \tfrac{d_{2}}{c_{2}}, G_{2} {=} \tfrac{\gamma_{2}}{c_{0}c_{2}}, R_{3} {=} \tfrac{-c_{0}\lambda_{2}}{F_{0}(c,d)}, C_{1} {=} \tfrac{-F_{0}(c,d)}{c_{0}^{2}\gamma_{3}}, L_{1} {=} \tfrac{c_{2}^{2}}{\gamma_{3}}.\\ \hline\end{array}$}
\label{fig:bq1c1l}
\end{figure}
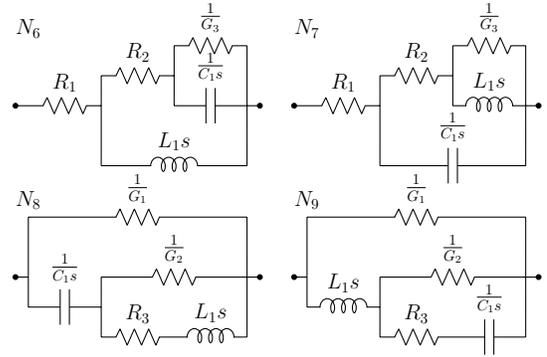

In \cite{JiangSmith11}, an algebraic description of the biquadratic impedances in $\mathcal{Z}_{2}$ was provided in terms of the polynomials
\begin{align}
c(s) &:= c_{2}s^{2} + c_{1}s + c_{0}, \text{ and} \nonumber\\
d(s) &:= d_{2}s^{2} + d_{1}s + d_{0}, \label{eq:cdd}
\end{align}
in a coprime factorisation $Z = c/d$ of the impedance $Z$. We summarize these results in the following lemma:
\begin{lemma}
\label{lem:bqrrc}
Let $Z = c/d$ where $c, d$ are as in (\ref{eq:cdd}); let $F_0(c,d) \neq 0$; and let $\gamma_1, \gamma_2, \gamma_3, \lambda_1, \lambda_2, \lambda_3$ and $\lambda_4$ be as in row (A1) of Table \ref{tab:term} (see the appendix).
Then $Z \in \mathcal{Z}_{2}$ if and only if at least one of the sets of constraints (Q1), (Q2), (Q3) or (Q4) in Table \ref{tab:cons} hold. Also, if $F_{0}(c,d) > 0$, then (Q1) (resp., (Q3)) holds if and only if (Q2) (resp., (Q4)) holds.

In addition,
\begin{enumerate}
\item If (Q1) holds and $F_{0}(c,d) < 0$, then $\gamma_{1} > 0$, $d_{0}, d_{2} \neq 0$, and $Z$ is the impedance of $N_{6}$ in Fig.\ \ref{fig:bq1c1l} when the element values are as indicated in that figure's caption.
\item If condition (Q3) holds and $F_{0}(c,d) < 0$, then $\gamma_{3} < 0$, $d_{0}, d_{2} {\neq} 0$, and $Z$ is the impedance of $N_{7}$ in Fig.\ \ref{fig:bq1c1l} when the element values are as indicated in that figure's caption.
\item If condition (Q4) holds and $F_{0}(c,d) < 0$, then $\gamma_{1} < 0$, $c_{0}, c_{2} \neq 0$, and $Z$ is the impedance of $N_{8}$ in Fig.\ \ref{fig:bq1c1l} when the element values are as indicated in that figure's caption.
\item If condition (Q2) holds and $F_{0}(c,d) < 0$, then $\gamma_{3} > 0$, $c_{0}, c_{2} \neq 0$, and $Z$ is the impedance of $N_{9}$ in Fig.\ \ref{fig:bq1c1l} when the element values are as indicated in that figure's caption.
\item If condition (Q1) holds and $F_{0}(c,d) > 0$, then $\gamma_{1} > 0$, $d_{0}, \lambda_{1} {\neq} 0$, and $Z$ is the impedance of $N_{5}$ in Fig. \ref{fig:bq2co2i} when the element values are as indicated in that figure's caption.
\item If condition (Q3) holds and $F_{0}(c,d) > 0$, then $\gamma_{3} < 0$, $d_{2}, \lambda_{3} {\neq} 0$, and $Z$ is the impedance of $N_{4}$ in Fig. \ref{fig:bq2co2i} when the element values are as indicated in that figure's caption.
\end{enumerate}
\end{lemma}
\begin{remark}
Note that conditions (Q1)--(Q4) in the above lemma are equivalent to $Z$ being \emph{regular} (i.e., $Z$ is positive-real and the least value of the real part of either $Z(j\omega)$ or $1/Z(j\omega)$ occurs at either $\omega = 0$ or $\omega = \infty$) \cite{JiangSmith11}.
\end{remark}

\begin{remark}
\label{rem:z2lo2c}
If $Z$ is the impedance of one of the networks $N_{4}$--$N_{9}$ in Figs.\ \ref{fig:bq2co2i}--\ref{fig:bq1c1l}, then it is straightforward to show that $Z$ is biquadratic. Also, if $c, d$ as in (\ref{eq:cdd}) are a coprime factorization for $Z$ (i.e., $Z = c/d$ and $F_0(c,d) \neq 0$), then the element values in these networks are uniquely determined by the coefficients $c_{0}$--$c_{2}$ and $d_{0}$--$d_{2}$ in accordance with the expressions in the captions of Figs.\ \ref{fig:bq2co2i}--\ref{fig:bq1c1l}.
\end{remark}

From Lemmas \ref{lem:bqmgs} and \ref{lem:bqrrc}, we obtain the following two lemmas, which provide an alternative characterisation for the set $\mathcal{Z}_{2}$. This characterisation will be used in Section \ref{sec:mgsbc} to obtain a minimal generating set for $\mathcal{Z}_{3}$. The papers \cite{LIN, JiangSmith11} provide an indirect proof of these two lemmas. For completeness, we present a more direct algebraic proof.
\begin{lemma}
\label{lem:bqec}
Let $c, d$ be as in (\ref{eq:cdd}), let $\gamma_{1}$--$\gamma_{3}$ and $\lambda_{1}$--$\lambda_{4}$ be as in row (A1) of Table \ref{tab:term}, let $F_0(c,d) < 0$, and consider the following set of inequalities:
\begin{remunerate}
\labitem{(Q\Alph{muni})}{nl:bqca1} $c_{0}d_{0}, d_{2}\lambda_{1}, \gamma_{2}d_{0}d_{2} \geq 0$; $d_{0}, d_{2} \neq 0$; and $\gamma_{1} > 0$.
\labitem{(Q\Alph{muni})}{nl:bqca2} $c_{2}d_{2}, d_{0}\lambda_{3}, -\gamma_{2}d_{0}d_{2} \geq 0$; $d_{0}, d_{2} \neq 0$; and $\gamma_{3} < 0$.
\labitem{(Q\Alph{muni})}{nl:bqca3} $c_{0}d_{0}, c_{2}\lambda_{4}, -\gamma_{2}c_{0}c_{2} \geq 0$; $c_{0}, c_{2} \neq 0$; and $\gamma_{1} < 0$.
\labitem{(Q\Alph{muni})}{nl:bqca4} $c_{2}d_{2}, c_{0}\lambda_{2}, \gamma_{2}c_{0}c_{2} \geq 0$; $c_{0}, c_{2} \neq 0$; and $\gamma_{3} > 0$.
\end{remunerate}
With (Q1)--(Q4) as in Table \ref{tab:cons}, then (Q1) (resp., (Q2), (Q3), (Q4)) is satisfied if and only if \ref{nl:bqca1} (resp., \ref{nl:bqca4}, \ref{nl:bqca3}, \ref{nl:bqca2}) is satisfied.
\end{lemma}

\begin{IEEEproof}
To show that (Q1) $\Rightarrow$ \ref{nl:bqca1}, it suffices to show that condition (Q1) and $F_{0}(c,d) < 0$ together imply that $d_{0}, d_{2} \neq 0$ and $\gamma_{1} > 0$. To see this, we note the following relationships:\footnote{Note that many of the relationships in this proof can be inferred from inspection of the Sylvester matrix $\mathcal{S}_{4}(c,d)$. For example, equation (\ref{eq:bql1c3}) follows from the expansion for $\abs{\mathcal{S}_{4}(c,d)}$ along the first column of $\mathcal{S}_{4}(c,d)$, noting that $\lambda_4$ (resp., $\lambda_1$) is the minor formed from columns $2, 3$ and $4$ and rows $2, 3$ and $4$ (resp., rows $1, 3$ and $4$) of $\mathcal{S}_{4}(c,d)$.}
\begin{align}
\lambda_{1} + d_{0}\gamma_{2} &= d_{1}\gamma_{1}, \label{eq:bql1c1} \\
\gamma_{2} + c_{0}d_{2} &= c_{2}d_{0}, \text{ and}  \label{eq:bql1c2} \\
c_{2}\lambda_{1} - F_{0}(c,d) &= -d_{2}\lambda_{4}. \label{eq:bql1c3}
\end{align}
Since $c_2$ and $\lambda_1$ have the same sign and $F_{0}(c,d) < 0$, then (\ref{eq:bql1c3}) implies that $d_2 \lambda_4 < 0$, so $d_{2} \neq 0$. If $d_{0} = 0$, then it follows from (\ref{eq:bql1c2}) that $c_{0}  = 0$ (since $d_{2} \neq 0$), but this again implies that $F_{0}(c,d) = 0$. We conclude that $d_{0}, d_{2} \neq 0$. Furthermore, if $\gamma_{1} \leq 0$, then it follows from (\ref{eq:bql1c1}) that $d_{1} = \gamma_{2} = \lambda_{1} = 0$ (since $d_{0} \neq 0$). But this implies that $F_{0}(c,d) = -c_{1}d_{2}\gamma_{1}$. Since $F_{0}(c,d) < 0$, then we conclude that $\gamma_{1}$ must be positive. It follows that $d_{0}, d_{2} \neq 0$ and $\gamma_{1} > 0$, which completes the proof of (Q1) $\Rightarrow$ \ref{nl:bqca1}.

To show that \ref{nl:bqca1} $\Rightarrow$ (Q1), we recall that (\ref{eq:bql1c1})--(\ref{eq:bql1c3}) hold, and we note the following additional relationships:
\begin{align}
\gamma_{2}d_{2}\lambda_{1} - F_{0}(c,d)d_{0}d_{2} &= d_{2}^{2}\gamma_{1}, \text{ and} \label{eq:bql1c4}\\
\gamma_{1} + c_{0}d_{1} &= c_{1}d_{0}. \label{eq:bql1c5}
\end{align}
Here, (\ref{eq:bql1c4}) implies that $d_{0}$ and $d_{2}$ have the same sign. Since, in addition, $d_{0}, d_{2} \neq 0$ and $\gamma_{2}d_{0}d_{2} \geq 0$, then we conclude that $\gamma_{2} \geq 0$. Then, (\ref{eq:bql1c2}) implies that $c_{2}$ and $d_{0}$ have the same sign, and so $c_{0}, c_{2}, d_{0}, d_{2}$ and $\lambda_{1}$ have the same sign, and $\gamma_{2} \geq 0$. Next, note that (\ref{eq:bql1c1}) implies that $d_{1}$ and $\lambda_{1}$ have the same sign. Finally, (\ref{eq:bql1c5}) implies that $c_{1}$ and $d_{0}$ have the same sign, which completes the proof of \ref{nl:bqca1} $\Rightarrow$ (Q1).

The proof of the remaining conditions are analogous. Specifically, in the above argument, we swap $c_{0}$ with $c_2$ and $d_0$ with $d_2$ to prove that (Q4) $\iff$ \ref{nl:bqca2}; we swap $c$ with $d$ to prove that (Q3) $\iff$ \ref{nl:bqca3}; and we swap $c_0$ with $d_2$, $c_1$ with $d_1$, and $c_2$ with $d_0$ to prove that (Q2) $\iff$ \ref{nl:bqca4}. 
\end{IEEEproof}

\begin{lemma}
Let $c, d$ be as in (\ref{eq:cdd}), let $\gamma_{1}$--$\gamma_{3}$ and $\lambda_{1}$--$\lambda_{4}$ be as in row (A1) of Table \ref{tab:term}, let $F_{0}(c,d) > 0$, and consider the following sets of inequalities:
\begin{remunerate}
\setcounter{muni}{4}
\labitem{(Q\Alph{muni})}{nl:bqca5} $c_{2}, d_{0}, d_{2}, \lambda_{3}$ have the same sign; $d_{2}, \lambda_{3} {\neq} 0$; and $\gamma_{3} {<} 0$.
\labitem{(Q\Alph{muni})}{nl:bqca6} $c_{0}, d_{0}, d_{2}, \lambda_{1}$ have the same sign; $d_{0}, \lambda_{1} {\neq} 0$; and $\gamma_{1} {>} 0$.
\end{remunerate}
The following hold:
\begin{enumerate}
\item Conditions (Q3), (Q4) and \ref{nl:bqca5} are all equivalent.\label{nl:bq1retc1}
\item Conditions (Q1), (Q2) and \ref{nl:bqca6} are all equivalent.\label{nl:bq1retc2}
\end{enumerate}
\end{lemma}

\begin{IEEEproof}
We first prove condition \ref{nl:bq1retc1}.

To show that (Q3) $\Rightarrow$ \ref{nl:bqca5}, it suffices to show that if (Q3) holds and $F_{0}(c,d) > 0$, then $d_{2}, \lambda_{3} \neq 0$ and $\gamma_{3} < 0$. To show this, we note the following relationships
\begin{align}
F_{0}(c,d)d_{2} + d_{0}\gamma_{3}^{2} &= -\gamma_{2}\lambda_{3}, \label{eq:bql1ca1} \\
\lambda_{3} + c_{2}d_{1}^{2} &= d_{2}(c_{2}d_{0} - c_{0}d_{2} + c_{1}d_{1}), \label{eq:bql1ca2} \\
\gamma_{2}^{2} + F_{0}(c,d) &= \gamma_{1}\gamma_{3}, \text{ and} \label{eq:bql1ca3} \\
\lambda_{3} - d_{2}\gamma_{2} &= -d_{1}\gamma_{3}. \label{eq:bql1ca4}
\end{align}
Since $F_{0}(c,d)>0$, then (\ref{eq:bql1ca3}) implies that $\gamma_1 \gamma_{3} > 0$, so $\gamma_{3} \neq 0$. Then, from (\ref{eq:bql1ca2}), $d_{2} = 0$ implies $c_{2} = 0$ or $d_{1} = 0$, and in either case we have $\gamma_{3} = 0$, so we conclude that $d_{2}, \gamma_{3} \neq 0$. Next, from (\ref{eq:bql1ca1}), we find that $\lambda_{3} = 0$ implies that either $F_{0}(c,d) = 0$ or $d_{2} = 0$, neither of which is possible, and so $d_{2}, \gamma_{3}, \lambda_{3} \neq 0$. Finally, since $d_{1}, d_{2}$ and $\lambda_{3}$ have the same sign and $\gamma_{2} \leq 0$, then (\ref{eq:bql1ca4}) implies that $\gamma_{3} < 0$.

To see that \ref{nl:bqca5} $\Rightarrow$ (Q3), we recall that (\ref{eq:bql1ca1})--(\ref{eq:bql1ca4}) hold, and we note the following additional relationships:
\begin{align}
\gamma_{3}^{2}d_{0}d_{2} + \lambda_{3}^{2} + F_{0}(c,d)d_{2}^{2} &= -d_{1}\gamma_{3}\lambda_{3}, \label{eq:bql1ca5} \\
-\gamma_{3} + c_{2}d_{1} &= c_{1}d_{2}, \text{ and} \label{eq:bql1ca6} \\
-\gamma_{1} + c_{1}d_{0} &= c_{0}d_{1}. \label{eq:bql1ca7}
\end{align}
Then (\ref{eq:bql1ca5}) implies that $d_{1}$ has the same sign as $\lambda_{3}$, whereupon (\ref{eq:bql1ca6}) implies that $c_{1}$ has the same sign as $d_{2}$. Also, (\ref{eq:bql1ca3}) implies that $\gamma_{1} \leq 0$, whereupon (\ref{eq:bql1ca7}) implies that $c_{0}$ has the same sign as $d_{1}$. Finally, (\ref{eq:bql1ca1}) implies that $\gamma_{2} \leq 0$.

An analogous argument proves that (Q4) is equivalent to:
\begin{remunerate}
\setcounter{muni}{6}
\labitem{(Q\Alph{muni})}{nl:bqca7} $c_{0}, c_{2}, d_{0}, \lambda_{4}$ have the same sign; $c_{0}, \lambda_{4} {\neq} 0$; and $\gamma_{1} {<} 0$.
\end{remunerate}
Accordingly, to complete the proof of the present lemma, we will show that \ref{nl:bqca5} and \ref{nl:bqca7} are equivalent. To see this, assume initially that \ref{nl:bqca5} holds and $d_{2} > 0$. Then $c_{2}, d_{0} \geq 0$ and the Sylvester matrix $\mathcal{S}_{4}(c,d)$ is positive definite. It follows that all of the principal minors of $\mathcal{S}_{4}(c,d)$ are positive, whence $c_{0} > 0$, $\gamma_{1} < 0$ and $\lambda_{4} > 0$. If, on the other hand, $d_{2} < 0$, then $\mathcal{S}_{4}(c,d)$ is negative definite, in which case $c_{0}, \gamma_{1}, \lambda_{4} < 0$. In either case, condition \ref{nl:bqca7} holds. A similar argument proves that, if \ref{nl:bqca7} holds and $F_{0}(c,d) > 0$, then \ref{nl:bqca5} holds.

The proof of condition \ref{nl:bq1retc2} is analogous to the above.
\end{IEEEproof}

\tikzset{ eqn1/.pic={
\ctikzset{bipoles/thickness=1}
\node[anchor=center] at (1,0) (O) {};
\draw (O) to[short,*-] ++(0.3,0)
to[R=$R_{1}$] ++(1.7,0)
to[short] ++(0,0.7)
to[/tikz/circuitikz/bipoles/length=1.65cm,R=$\frac{1}{G_{1}}$] ++(1.7,0)
to[short] ++(0,-0.7)
to[short,-*] ++(0.3,0)
($(O)+(2.0,0)$) to[short] ++(0,-0.7)
to[generic=$Z_{1}(s)$] ++(1.7,0)
to[short] ++(0,0.7);}}

\tikzset{ eqn2/.pic={
\ctikzset{bipoles/thickness=1}
\node[anchor=center] at (1,0) (O) {};
\draw (O) to[short,*-] ++(0.3,0)
to[short] ++(0,0.7)
to[R=$\frac{1}{G_{2}}$] ++(3.4,0)
to[short] ++(0,-0.7)
to[short,-*] ++(0.3,0)
($(O)+(0.3,0)$) to[short] ++(0,-0.7)
to[/tikz/circuitikz/bipoles/length=1.65cm,R=$R_{2}$] ++(1.7,0)
to[generic=$Z_{2}(s)$] ++(1.7,0)
to[short] ++(0,0.7);}}

\section{Non-minimal generating sets for $\mathcal{Z}_{1,2}$ and $\mathcal{Z}_{2,1}$}
\label{sec:nmgs}

The main contribution of this section is to derive generating sets for the classes $\mathcal{Z}_{1,2}$ and $\mathcal{Z}_{2,1}$. These are described in the following two theorems:

\begin{theorem}
\label{thm:z12nmgs}
The union of $\mathcal{N}_{1}$--$\mathcal{N}_{3}$, $\mathcal{N}_{5}$--$\mathcal{N}_{9}$, $\mathcal{N}_{10}$--$\mathcal{N}_{14}$ and $\mathcal{N}_{10}^{p}$--$\mathcal{N}_{14}^{p}$ is a generating set for $\mathcal{Z}_{1,2}$ (see Figs.\ \ref{fig:c0r}, \ref{fig:bl1in}, and \ref{fig:bq2co2i}--\ref{fig:bcr6}).
\end{theorem}

\begin{theorem}
\label{thm:z21nmgs}
The union of $\mathcal{N}_{1}$--$\mathcal{N}_{4}$, $\mathcal{N}_{6}$--$\mathcal{N}_{9}$, $\mathcal{N}_{10}^{i}$--$\mathcal{N}_{14}^{i}$ and $\mathcal{N}_{10}^{d}$--$\mathcal{N}_{14}^{d}$ is a generating set for $\mathcal{Z}_{2,1}$.
\end{theorem}

Note from \cite[Corollary 2]{HMS_GEN} that the network classes $\mathcal{N}_{10}$--$\mathcal{N}_{14}$ are not generic (and neither are the network classes $\mathcal{N}_{10}^{i}$, $\mathcal{N}_{10}^{d}$, $\mathcal{N}_{10}^{di}$, and so forth.).

We will prove Theorem \ref{thm:z12nmgs}, and Theorem \ref{thm:z21nmgs} can be proved in a similar manner. Our proof relies on the following well known network transformation:
\begin{lemma}
\label{lem:eqr}
For an arbitrary given impedance $H \in \mathbb{R}(s)$, the two networks in Fig.\ \ref{fig:eqn} have equivalent impedance under the transformations $R_{2} = R_{1}(1+R_{1}G_{1})$, $G_{2} = G_{1}/(1+R_{1}G_{1})$, and $\phi_{2} = \phi_{1}(1+R_{1}G_{1})^{2}$ (equivalently, $R_{1} = R_{2}/(1+R_{2}G_{2})$, $G_{1} = G_{2}(1+R_{2}G_{2})$, and $\phi_{1} = \phi_{2}/(1+R_{2}G_{2})^{2}$).
\end{lemma}

\tikzset{ eqn1/.pic={
\ctikzset{bipoles/thickness=1}
\node[anchor=center] at (1,0) (O) {};
\draw (O) to[short,*-] ++(0.3,0)
to[R=$R_{1}$] ++(1.7,0)
to[short] ++(0,0.9)
to[/tikz/circuitikz/bipoles/length=1.65cm,R=$\frac{1}{G_{1}}$] ++(2.0,0)
to[short] ++(0,-0.9)
to[short,-*] ++(0.3,0)
($(O)+(2.0,0)$) to[short] ++(0,-0.9)
to[twoport, t={\Large $N_{1}$}, l=$\phi_{1}H(s)$] ++(2.0,0)
to[short] ++(0,0.9);}}

\tikzset{ eqn2/.pic={
\ctikzset{bipoles/thickness=1}
\node[anchor=center] at (1,0) (O) {};
\draw (O) to[short,*-] ++(0.3,0)
to[short] ++(0,0.9)
to[R=$\frac{1}{G_{2}}$] ++(3.4,0)
to[short] ++(0,-0.9)
to[short,-*] ++(0.3,0)
($(O)+(0.3,0)$) to[short] ++(0,-0.9)
to[/tikz/circuitikz/bipoles/length=1.65cm,R=$R_{2}$] ++(1.7,0)
to[twoport, t={\Large $N_{2}$}, l=$\phi_{2}H(s)$] ++(1.7,0)
to[short] ++(0,0.9);}}

\begin{figure}[!t]
\centering
\normalsize
\begingroup
\renewcommand*{\arraystretch}{1.7}
\begin{tikzpicture}[scale=0.57, every node/.style={transform shape}, font=\Large]
\pic at (0,0) [] {eqn1};
\pic at (5.0,0) [] {eqn2};
\end{tikzpicture}
\endgroup
\caption{Two networks with equivalent impedance.}
\label{fig:eqn}
\end{figure}
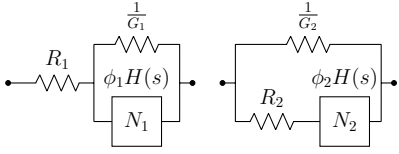

\begin{IEEEproof}[Proof of Theorem \ref{thm:z12nmgs}]
The proof of this theorem is similar to the method proposed in \cite[Section 2(b)]{ZJ_SIA}. In contrast to that paper, we will use the network transformation in Lemma \ref{lem:eqr} to eliminate several redundant elements.

If $Z$ is the impedance of a series-parallel network containing two or fewer energy storage elements, then $Z$ is the impedance of a network from one of the classes $\mathcal{N}_{1}$--$\mathcal{N}_{3}$ or $\mathcal{N}_{5}$--$\mathcal{N}_{9}$. Accordingly, it remains to consider the case in which $Z \in \mathcal{Z}_{1,2}$ is the impedance of a series-parallel network $N$ that contains exactly three energy storage elements. Then, at some stage in the construction of $N$, a network $N_{a}$ containing one energy storage element is connected either in series or parallel with a network $N_{b}$ containing two energy storage elements, and all subsequent stages in the construction of $N$ involve the addition of resistors in series or in parallel. 

Consider first the case in which $N_{a}$ and $N_{b}$ are connected in parallel. Since a series or parallel connection of two resistors can always be realized by a single resistor, then it is easily shown from Lemma \ref{lem:eqr} that $Z$ is realized by a network of the form of $N_{u}$ in Fig.\ \ref{fig:gsz12}. Since $N_{a}$ contains one energy storage element, then its impedance is realized by a network from $\mathcal{N}_{2}$ or $\mathcal{N}_{3}$, and by a network from $\mathcal{N}_{2a}$ or $\mathcal{N}_{3a}$, by Lemma \ref{lem:blgs}. It follows that the impedance of a network of the form of $N_{u}$ in Fig.\ \ref{fig:gsz12} can be realized by the impedance of a network of the form of $N_{v}$ or $N_{w}$ in that figure, where $N_{c}$ is a series-parallel network containing at most two energy storage elements. Furthermore, in network $N_{v}$ (resp., $N_{w}$), the network $N_{c}$ necessarily contains at most one inductor and one capacitor (resp., at most two inductors and no capacitors). Thus, from Section \ref{sec:blbqi}, it follows that $Z$ is the impedance of a network from one of the classes $\mathcal{N}_{2}$, $\mathcal{N}_{3}$, $\mathcal{N}_{5}$--$\mathcal{N}_{9}$, or $\mathcal{N}_{10}$--$\mathcal{N}_{14}$.

The case with $N_{a}$ and $N_{b}$ connected in series is similar. In this case, we find that $Z$ is the impedance of a network from one of the classes $\mathcal{N}_{2}$, $\mathcal{N}_{3}$, $\mathcal{N}_{5}$--$\mathcal{N}_{9}$, or $\mathcal{N}_{10}^{p}$--$\mathcal{N}_{14}^{p}$
\end{IEEEproof}

\tikzset{ gn1/.pic={
\ctikzset{bipoles/thickness=1}
\node[anchor=center] at (1,0) (O) {};
\draw (O) to[short,*-] ++(0.3,0)
to[R] ++(1.7,0)
to[short] ++(0,1.4)
to[twoport, t={\Large $N_{a}$}] ++(1.7,0)
to[short] ++(0,-1.4)
($(O)+(2.0,0)$) to[twoport, t={\Large $N_{b}$}] ++(1.7,0)
to[short,-*] ++(0.3,0)
($(O)+(2.0,0)$) to[short] ++(0,-1.4)
to[R] ++(1.7,0)
to[short] ++(0,1.4);}}

\tikzset{ gn2/.pic={
\ctikzset{bipoles/thickness=1}
\node[anchor=center] at (1,0) (O) {};
\draw (O) to[short,*-] ++(0.3,0)
to[R] ++(1.7,0)
to[short] ++(0,0.7)
to[R] ++(1.7,0)
to[L] ++(1.7,0)
to[short] ++(0,-0.7)
to[short,-*] ++(0.3,0)
($(O)+(2.0,0)$) to[short] ++(0,-0.7)
to[twoport, t={\Large $N_{c}$}] ++(3.4,0)
to[short] ++(0,0.7);}}

\tikzset{ gn3/.pic={
\ctikzset{bipoles/thickness=1}
\node[anchor=center] at (1,0) (O) {};
\draw (O) to[short,*-] ++(0.3,0)
to[R] ++(1.7,0)
to[short] ++(0,0.7)
to[R] ++(1.7,0)
to[C] ++(1.7,0)
to[short] ++(0,-0.7)
to[short,-*] ++(0.3,0)
($(O)+(2.0,0)$) to[short] ++(0,-0.7)
to[twoport, t={\Large $N_{c}$}] ++(3.4,0)
to[short] ++(0,0.7);}}

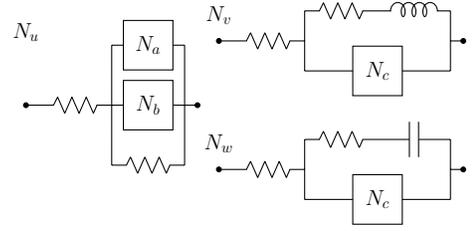
\begin{figure}[!t]
\centering
\normalsize
\begingroup
\renewcommand*{\arraystretch}{1.7}
\begin{tikzpicture}[scale=0.57, every node/.style={transform shape}, font=\Large]
\node at (1.0,1.7) {$N_{u}$};
\pic at (0,0) [] {gn1};
\node at (5.5,2.1) {$N_{v}$};
\pic at (4.5,1.5) [] {gn2};
\node at (5.5,-0.9) {$N_{w}$};
\pic at (4.5,-1.5) [] {gn3};
\end{tikzpicture}
\endgroup
\caption{Structure of the generating sets for $\mathcal{Z}_{1,2}$ in Theorem \ref{thm:z12nmgs}.}
\label{fig:gsz12}
\end{figure}

\tikzset{ clbc1a/.pic={
\ctikzset{bipoles/thickness=1}
\node[anchor=center] at (1,0) (O) {};
\draw (O) to[short,*-] ++(0.3,0)
to[R=$R_{4}$] ++(1.7,0)
to[short] ++(0,-1.4)
to[R=$R_{1}$] ++(1.7,0)
to[short] ++(0,0.7)
to[R=$R_{2}$] ++(1.7,0)
to[short] ++(0,0.7)
to[R=$\frac{1}{G_{3}}$] ++(1.7,0)
to[short,-*] ++(0.3,0)
($(O)+(5.4,-0.7)$) to[short] ++(0,-0.7)
to[C=$\frac{1}{C_{1}s}$] ++(1.7,0)
to[short] ++(0,1.4)
($(O)+(3.7,-1.4)$) to[short] ++(0,-1.4)
to[L=$L_{1}s$] ++(3.4,0)
to[short] ++(0,1.4)
($(O)+(2.0,0)$) to[short] ++(0,1.4)
to[R=$R_{5}$] ++(2.55,0)
to[L=$L_{2}s$] ++(2.55,0)
to[short] ++(0,-1.4);}}

\tikzset{ clbc1b/.pic={
\ctikzset{bipoles/thickness=1}
\node[anchor=center] at (1,0) (O) {};
\draw (O) to[short,*-] ++(0.3,0)
to[R=$R_{4}$] ++(1.7,0)
to[short] ++(0,-1.4)
to[R=$R_{1}$] ++(1.7,0)
to[short] ++(0,0.7)
to[R=$R_{2}$] ++(1.7,0)
to[short] ++(0,0.7)
to[R=$\frac{1}{G_{3}}$] ++(1.7,0)
to[short,-*] ++(0.3,0)
($(O)+(5.4,-0.7)$) to[short] ++(0,-0.7)
to[L=$L_{1}s$] ++(1.7,0)
to[short] ++(0,1.4)
($(O)+(3.7,-1.4)$) to[short] ++(0,-1.4)
to[C=$\frac{1}{C_{1}s}$] ++(3.4,0)
to[short] ++(0,1.4)
($(O)+(2.0,0)$) to[short] ++(0,1.4)
to[R=$R_{5}$] ++(2.55,0)
to[L=$L_{2}s$] ++(2.55,0)
to[short] ++(0,-1.4);}}

\begin{figure}[!t]
\centering
\normalsize
\begin{tikzpicture}[scale=0.57, every node/.style={transform shape}, font=\Large]
\pic at (0,0) [] {clbc1a};
\pic at (8.0,0) [] {clbc1b};
\node at (1.5,1.5) {$N_{10}$};
\node at (9.5,1.5) {$N_{11}$};
\end{tikzpicture}
\caption[]{Networks $N_{10}$ and $N_{11}$. Here, (i) $R_{1}, R_{2}, R_{4}, R_{5} \geq 0$; (ii) $G_{3} \geq 0$; and (iii) $C_{1}, L_{1}, L_{2} > 0$. We define $\mathcal{N}_{10}$ (resp., $\mathcal{N}_{11}$) as the set of all networks of the form of $N_{10}$ (resp., $N_{11}$) that satisfy conditions (i)--(iii). 

\quad If the relevant conditions of Theorem \ref{thm:bcn1013} hold, and $\alpha , \beta$, $c_0$--$c_2$, $d_0$--$d_2$, $\kappa$, $\gamma_1$--$\gamma_3$ and $\lambda_1$--$\lambda_4$ are as defined in that theorem, then $N_{10}$ and $N_{11}$ have impedance $a/b$ with $a, b \in \mathbb{R}[s]$ as in (\ref{eq:abd}) when the element values are: \\[0.2em] $\begin{array}{|l|l|}\hline \rule{0pt}{1.1\normalbaselineskip} N_{10} & R_{1} {=} \tfrac{\kappa (\tilde{x})c_{0}(\tilde{x})}{d_{0}(\tilde{x})}, R_{2} {=} \tfrac{\kappa (\tilde{x})\gamma_{2}(\tilde{x})}{d_{0}(\tilde{x})d_{2}(\tilde{x})}, G_{3} {=} \tfrac{d_{2}(\tilde{x})\lambda_{1}(\tilde{x})}{\beta ^{4}(\tilde{x})(\kappa (\tilde{x}))^{2}F_0(a,b)}, \\ & R_{4} {=} \tfrac{\alpha (\tilde{x})}{\beta (\tilde{x})}, R_{5} {=} \tfrac{\kappa (\tilde{x})\tilde{x}}{(\beta (\tilde{x}))^{2}}, C_{1} {=} \tfrac{(d_{2}(\tilde{x}))^{2}\gamma_{1}(\tilde{x})}{(\beta (\tilde{x}))^{4}(\kappa (\tilde{x}))^{2}F_0(a,b)}, \\ & L_{1} {=} \tfrac{\kappa (\tilde{x})\gamma_{1}(\tilde{x})}{(d_{0}(\tilde{x}))^{2}}, L_{2} {=} \tfrac{\kappa (\tilde{x})}{(\beta (\tilde{x}))^{2}}. \\ \hline \rule{0pt}{1.1\normalbaselineskip} N_{11} & R_{1} {=} \tfrac{\kappa (\tilde{x})c_{2}(\tilde{x})}{d_{2}(\tilde{x})}, R_{2} {=} \tfrac{-\kappa (\tilde{x})\gamma_{2}(\tilde{x})}{d_{0}(\tilde{x})d_{2}(\tilde{x})}, G_{3} {=} \tfrac{d_{0}(\tilde{x})\lambda_{3}(\tilde{x})}{(\beta (\tilde{x}))^{4}(\kappa (\tilde{x}))^{2}F_0(a,b)}, \\
&R_{4} {=} \tfrac{\alpha (\tilde{x})}{\beta (\tilde{x})}, R_{5} {=} \tfrac{\kappa (\tilde{x})\tilde{x}}{(\beta (\tilde{x}))^{2}}, C_{1} {=} \tfrac{-(d_{2}(\tilde{x}))^{2}}{\kappa (\tilde{x})\gamma_{3}(\tilde{x})}, \\
&L_{1} {=} \tfrac{-(\beta (\tilde{x}))^{4}(\kappa (\tilde{x}))^{2}F_0(a,b)}{(d_{0}(\tilde{x}))^{2}\gamma_{3}(\tilde{x})}, L_{2} {=} \tfrac{\kappa (\tilde{x})}{(\beta (\tilde{x}))^{2}}. \\ \hline\end{array}$ \\[0.2em]

\quad If the relevant conditions of Theorem \ref{thm:bqn1013} hold, and $\alpha, \beta $, $c_0$--$c_2$, $d_0$--$d_2$, $\eta$, $\gamma_1$--$\gamma_3$ and $\lambda_1$--$\lambda_4$ are as defined in that theorem, then $N_{10}$ and $N_{11}$ have impedance $a/b$ with $a, b \in \mathbb{R}[s]$ as in (\ref{eq:abd2}) when the element values are: \\[0.2em]
$\begin{array}{|l|l|}\hline \rule{0pt}{1.1\normalbaselineskip} N_{10} & R_{1} {=} \tfrac{c_{0}(\tilde{x})}{d_{0}(\tilde{x},\tilde{z})}, R_{2} {=} \tfrac{\gamma_{2}(\tilde{x},\tilde{z})}{d_{0}(\tilde{x},\tilde{z})d_{2}(\tilde{x},\tilde{z})}, \\ &G_{3} {=} \tfrac{-d_{2}(\tilde{x},\tilde{z})\lambda_{1}(\tilde{x},\tilde{z})}{(\beta (\tilde{x}))^{3}\eta (\tilde{x},\tilde{z})F_0(a,b)}, R_{4} {=} \tfrac{\alpha (\tilde{x})}{\beta (\tilde{x})}, R_{5} {=} \tfrac{\tilde{x}}{\beta (\tilde{x})\tilde{z}}, \\ &C_{1} {=} \tfrac{-(d_{2}(\tilde{x},\tilde{z}))^{2}\gamma_{1}(\tilde{x},\tilde{z})}{(\beta (\tilde{x}))^{3}\eta (\tilde{x},\tilde{z})F_0(a,b)}, L_{1} {=} \tfrac{\gamma_{1}(\tilde{x},\tilde{z})}{(d_{0}(\tilde{x},\tilde{z}))^{2}}, L_{2} {=} \tfrac{1}{\beta (\tilde{x})\tilde{z}}. \\ \hline \rule{0pt}{1.1\normalbaselineskip} N_{11} & R_{1} {=} \tfrac{c_{2}(\tilde{x})}{d_{2}(\tilde{x},\tilde{z})}, R_{2} {=} \tfrac{-\gamma_{2}(\tilde{x},\tilde{z})}{d_{0}(\tilde{x},\tilde{z})d_{2}(\tilde{x},\tilde{z})}, \\ &G_{3} {=} \tfrac{-d_{0}(\tilde{x},\tilde{z})\lambda_{3}(\tilde{x},\tilde{z})}{(\beta (\tilde{x}))^{3}\eta (\tilde{x},\tilde{z})F_0(a,b)}, R_{4} {=} \tfrac{\alpha (\tilde{x})}{\beta (\tilde{x})}, R_{5} {=} \tfrac{\tilde{x}}{\beta (\tilde{x})\tilde{z}}, \\ &C_{1} {=} \tfrac{-(d_{2}(\tilde{x},\tilde{z}))^{2}}{\gamma_{3}(\tilde{x},\tilde{z})}, L_{1} {=} \tfrac{(\beta (\tilde{x}))^{3}\eta (\tilde{x},\tilde{z})F_0(a,b)}{(d_{0}(\tilde{x},\tilde{z}))^{2}\gamma_{3}(\tilde{x},\tilde{z})}, L_{2} {=} \tfrac{1}{\beta (\tilde{x})\tilde{z}}. \\ \hline\end{array}$}
\label{fig:bcr2}
\end{figure}

\tikzset{ clbc1c/.pic={
\ctikzset{bipoles/thickness=1}
\node[anchor=center] at (1,0) (O) {};
\draw (O) to[short,*-] ++(0.3,0)
to[R=$R_{4}$] ++(1.7,0)
to[short] ++(0,-2.1)
to[C=$\frac{1}{C_{1}s}$] ++(1.7,0)
to[short] ++(0,0.7)
to[R=$\frac{1}{G_{2}}$] ++(3.4,0)
to[short] ++(0,1.4)
($(O)+(3.7,-2.1)$) to[short] ++(0,-0.7)
to[R=$R_{3}$] ++(1.7,0)
to[L=$L_{1}s$] ++(1.7,0)
to[short] ++(0,1.4)
($(O)+(2.0,0)$) to[R=$\frac{1}{G_{1}}$] ++(5.1,0)
to[short,-*] ++(0.3,0)
($(O)+(2.0,0)$) to[short] ++(0,1.4)
to[R=$R_{5}$] ++(2.55,0)
to[L=$L_{2}s$] ++(2.55,0)
to[short] ++(0,-1.4);}}

\tikzset{ clbc1d/.pic={
\ctikzset{bipoles/thickness=1}
\node[anchor=center] at (1,0) (O) {};
\draw (O) to[short,*-] ++(0.3,0)
to[R=$R_{4}$] ++(1.7,0)
to[short] ++(0,-2.1)
to[L=$L_{1}s$] ++(1.7,0)
to[short] ++(0,0.7)
to[R=$\frac{1}{G_{2}}$] ++(3.4,0)
to[short] ++(0,1.4)
($(O)+(3.7,-2.1)$) to[short] ++(0,-0.7)
to[R=$R_{3}$] ++(1.7,0)
to[C=$\frac{1}{C_{1}s}$] ++(1.7,0)
to[short] ++(0,1.4)
($(O)+(2.0,0)$) to[R=$\frac{1}{G_{1}}$] ++(5.1,0)
to[short,-*] ++(0.3,0)
($(O)+(2.0,0)$) to[short] ++(0,1.4)
to[R=$R_{5}$] ++(2.55,0)
to[L=$L_{2}s$] ++(2.55,0)
to[short] ++(0,-1.4);}}

\begin{figure}[!t]
\centering
\normalsize
\begin{tikzpicture}[scale=0.57, every node/.style={transform shape}, font=\Large]
\pic at (0,0) [] {clbc1c};
\pic at (8.0,0) [] {clbc1d};
\node at (1.5,1.5) {$N_{12}$};
\node at (9.5,1.5) {$N_{13}$};
\end{tikzpicture}
\caption[]{Networks $N_{12}$ and $N_{13}$. Here, (i) $R_{3}, R_{4}, R_{5} \geq 0$; (ii) $G_{1}, G_{2} \geq 0$; and (iii) $C_{1}, L_{1}, L_{2} > 0$. We define $\mathcal{N}_{12}$ (resp., $\mathcal{N}_{13}$) as the set of all networks of the form of $N_{12}$ (resp., $N_{13}$) that satisfy conditions (i)--(iii). 

\quad If the relevant conditions of Theorem \ref{thm:bcn1013} hold, and $\alpha , \beta$, $c_0$--$c_2$, $d_0$--$d_2$, $\kappa$, $\gamma_1$--$\gamma_3$ and $\lambda_1$--$\lambda_4$ are as defined in that theorem, then $N_{12}$ and $N_{13}$ have impedance $a/b$ with $a, b \in \mathbb{R}[s]$ as in (\ref{eq:abd}) when the element values are: \\[0.2em] $\begin{array}{|l|l|}\hline \rule{0pt}{1.1\normalbaselineskip} N_{12} & G_{1} {=} \tfrac{d_{0}(\tilde{x})}{\kappa (\tilde{x})c_{0}(\tilde{x})}, G_{2} {=} \tfrac{-\gamma_{2}(\tilde{x})}{\kappa (\tilde{x})c_{0}(\tilde{x})c_{2}(\tilde{x})}, R_{3} {=} \tfrac{c_{2}(\tilde{x})\lambda_{4}(\tilde{x})}{(\beta (\tilde{x}))^{4}F_0(a,b)}, \\ & R_{4} {=} \tfrac{\alpha (\tilde{x})}{\beta (\tilde{x})}, R_{5} {=} \tfrac{\kappa (\tilde{x})\tilde{x}}{(\beta (\tilde{x}))^{2}}, C_{1} {=} \tfrac{-\gamma_{1}(\tilde{x})}{\kappa (\tilde{x})(c_{0}(\tilde{x}))^{2}}, \\ & L_{1} {=} \tfrac{-(c_{2}(\tilde{x}))^{2}\gamma_{1}(\tilde{x})}{(\beta (\tilde{x}))^{4}F_0(a,b)}, L_{2} {=} \tfrac{\kappa (\tilde{x})}{(\beta (\tilde{x}))^{2}}. \\ \hline \rule{0pt}{1.1\normalbaselineskip} N_{13} & G_{1} {=} \tfrac{d_{2}(\tilde{x})}{\kappa (\tilde{x})c_{2}(\tilde{x})}, G_{2} {=} \tfrac{\gamma_{2}(\tilde{x})}{\kappa (\tilde{x})c_{0}(\tilde{x})c_{2}(\tilde{x})}, R_{3} {=} \tfrac{c_{0}(\tilde{x})\lambda_{2}(\tilde{x})}{(\beta (\tilde{x}))^{4}F_0(a,b)}, \\ & R_{4} {=} \tfrac{\alpha (\tilde{x})}{\beta (\tilde{x})}, R_{5} {=} \tfrac{\kappa (\tilde{x})\tilde{x}}{(\beta (\tilde{x}))^{2}}, C_{1} {=} \tfrac{(\beta (\tilde{x}))^{4}F_0(a,b)}{(c_{0}(\tilde{x}))^{2}\gamma_{3}(\tilde{x})}, \\ & L_{1} {=} \tfrac{\kappa (\tilde{x})(c_{2}(\tilde{x}))^{2}}{\gamma_{3}(\tilde{x})}, L_{2} {=} \tfrac{\kappa (\tilde{x})}{(\beta (\tilde{x}))^{2}}. \\ \hline\end{array}$ \\[0.2em]

\quad If the relevant conditions of Theorem \ref{thm:bqn1013} hold, and $\alpha, \beta $, $c_0$--$c_2$, $d_0$--$d_2$, $\eta$, $\gamma_1$--$\gamma_3$ and $\lambda_1$--$\lambda_4$ are as defined in that theorem, then $N_{12}$ and $N_{13}$ have impedance $a/b$ with $a, b \in \mathbb{R}[s]$ as in (\ref{eq:abd2}) when the element values are: \\[0.2em]
$\begin{array}{|l|l|}\hline \rule{0pt}{1.1\normalbaselineskip} N_{12} & G_{1} {=} \tfrac{d_{0}(\tilde{x},\tilde{z})}{c_{0}(\tilde{x})}, G_{2} {=} \tfrac{-\gamma_{2}(\tilde{x},\tilde{z})}{c_{0}(\tilde{x})c_{2}(\tilde{x})}, \\
&R_{3} {=} \tfrac{-c_{2}(\tilde{x})\lambda_{4}(\tilde{x},\tilde{z})}{(\beta (\tilde{x}))^{3}\eta (\tilde{x},\tilde{z})F_0(a,b)}, R_{4} {=} \tfrac{\alpha (\tilde{x})}{\beta (\tilde{x})}, R_{5} {=} \tfrac{\tilde{x}}{\beta (\tilde{x})\tilde{z}}, \\ &C_{1} {=} \tfrac{-\gamma_{1}(\tilde{x},\tilde{z})}{(c_{0}(\tilde{x},\tilde{z}))^{2}}, L_{1} {=} \tfrac{(c_{2}(\tilde{x}))^{2}\gamma_{1}(\tilde{x},\tilde{z})}{(\beta (\tilde{x}))^{3}\eta (\tilde{x},\tilde{z})F_0(a,b)}, L_{2} {=} \tfrac{1}{\beta (\tilde{x})\tilde{z}}. \\ \hline \rule{0pt}{1.1\normalbaselineskip} N_{13} & G_{1} {=} \tfrac{d_{2}(\tilde{x},\tilde{z})}{c_{2}(\tilde{x})}, G_{2} {=} \tfrac{\gamma_{2}(\tilde{x},\tilde{z})}{c_{0}(\tilde{x})c_{2}(\tilde{x})}, \\
&R_{3} {=} \tfrac{-c_{0}(\tilde{x})\lambda_{2}(\tilde{x},\tilde{z})}{(\beta (\tilde{x}))^{3}\eta (\tilde{x},\tilde{z})F_0(a,b)}, R_{4} {=} \tfrac{\alpha (\tilde{x})}{\beta (\tilde{x})}, R_{5} {=} \tfrac{\tilde{x}}{\beta (\tilde{x})\tilde{z}}, \\ &C_{1} {=} \tfrac{-(\beta (\tilde{x}))^{3}\eta (\tilde{x},\tilde{z})F_0(a,b)}{(c_{0}(\tilde{x}))^{2}\gamma_{3}(\tilde{x},\tilde{z})}, L_{1} {=} \tfrac{(c_{2}(\tilde{x}))^{2}}{\gamma_{3}(\tilde{x},\tilde{z})}, L_{2} {=} \tfrac{1}{\beta (\tilde{x})\tilde{z}}. \\ \hline\end{array}$}
\label{fig:bcr5}
\end{figure}

\tikzset{ clbc2a/.pic={
\ctikzset{bipoles/thickness=1}
\node[anchor=center] at (1,0) (O) {};
\draw (O) to[short,*-] ++(0.3,0)
to[R=$R_{4}$] ++(1.7,0)
to[short] ++(0,-1.4)
to[R=$R_{1}$] ++(1.7,0)
to[short] ++(0,0.7)
to[R=$R_{2}$] ++(1.7,0)
to[short] ++(0,0.7)
to[R=$\frac{1}{G_{3}}$] ++(1.7,0)
to[short,-*] ++(0.3,0)
($(O)+(5.4,-0.7)$) to[short] ++(0,-0.7)
to[L=$L_{1}s$] ++(1.7,0)
to[short] ++(0,1.4)
($(O)+(3.7,-1.4)$) to[short] ++(0,-1.4)
to[L=$L_{2}s$] ++(3.4,0)
to[short] ++(0,1.4)
($(O)+(2.0,0)$) to[short] ++(0,1.4)
to[R=$R_{5}$] ++(2.55,0)
to[C=$\frac{1}{C_{1}s}$] ++(2.55,0)
to[short] ++(0,-1.4);}}

\begin{figure}[!t]
\centering
\normalsize
\begin{tikzpicture}[scale=0.57, every node/.style={transform shape}, font=\Large]
\node at (-1.6,1.0) {$N_{14}$};
\node at (-1.6,-2.0) {$\begin{array}{rl}\text{(i)}& R_{1}, R_{4}, R_{5} \geq 0\\
\text{(ii)}& R_{2} > 0\\
\text{(iii)}& G_{3} \geq 0\\
\text{(iv)}& C_{1}, L_{1}, L_{2} > 0
\end{array}$};
\pic at (0,0) [] {clbc2a};
\end{tikzpicture}
\caption[]{Network $N_{14}$. We define $\mathcal{N}_{14}$ as the set of all networks of the form of $N_{14}$ that satisfy conditions (i)--(iv).

\quad If the relevant conditions of Theorem \ref{thm:bcn14} hold, and $\alpha , \beta$, $c_0$--$c_2$, $d_0$--$d_2$, $\kappa$, $\gamma_1$--$\gamma_3$ and $\lambda_1$--$\lambda_4$ are as defined in that theorem, then $N_{14}$ has impedance $a/b$ with $a, b \in \mathbb{R}[s]$ as in (\ref{eq:abd}) when the element values are: \\[0.2em]
$\begin{array}{|l|}\hline \rule{0pt}{1.1\normalbaselineskip} R_{1} {=} \tfrac{\kappa (\tilde{x})c_{0}(\tilde{x})}{d_{0}(\tilde{x})}, R_{2} {=} \tfrac{\kappa (\tilde{x})(\gamma_{1}(\tilde{x}))^{2}}{d_{0}(\tilde{x})\lambda_{1}(\tilde{x})}, G_{3} {=} \tfrac{d_{2}(\tilde{x})\lambda_{1}(\tilde{x})}{(\kappa (\tilde{x}))^{2}(\beta (\tilde{x}))^{4}F_0(a,b)}, \\
R_{4} {=} \tfrac{\alpha (\tilde{x})}{\beta (\tilde{x})}, R_{5} {=} \tfrac{\kappa (\tilde{x})\tilde{x}}{(\beta (\tilde{x}))^{2}}, C_{1} {=} \tfrac{(\beta (\tilde{x}))^{2}}{\kappa (\tilde{x})},\\
L_{1} {=} \tfrac{(\beta (\tilde{x}))^{4}(\kappa (\tilde{x}))^{2}\gamma_{1}(\tilde{x})F_0(a,b)}{(\lambda_{1}(\tilde{x}))^{2}}, L_{2} {=} \tfrac{\kappa (\tilde{x})\gamma_{1}(\tilde{x})}{(d_{0}(\tilde{x}))^{2}}. \\ \hline\end{array}$ \\[0.2em]

\quad If the relevant conditions of Theorem \ref{thm:bqn14} hold, and $\alpha, \beta$, $c_0$--$c_2$ $d_0$--$d_2$, $\eta$, $\gamma_1$--$\gamma_3$ and $\lambda_1$--$\lambda_4$ are as defined in that theorem, then $N_{14}$ has impedance $a/b$ with $a, b \in \mathbb{R}[s]$ as in (\ref{eq:abd2}) when the element values are:  \\[0.2em]
$\begin{array}{|l|}\hline \rule{0pt}{1.1\normalbaselineskip} R_{1} {=} \tfrac{c_{0}(\tilde{x})}{d_{0}(\tilde{x},\tilde{z})}, R_{2} {=} \tfrac{(\gamma_{1}(\tilde{x},\tilde{z}))^{2}}{d_{0}(\tilde{x},\tilde{z})\lambda_{1}(\tilde{x},\tilde{z})}, G_{3} {=} \tfrac{d_{2}(\tilde{x},\tilde{z})\lambda_{1}(\tilde{x},\tilde{z})}{(\beta (\tilde{x}))^{3}\eta (\tilde{x},\tilde{z})F_0(a,b)}, \\
R_{4} {=} \tfrac{\alpha (\tilde{x})}{\beta (\tilde{x})}, R_{5} {=} \tfrac{\tilde{x}}{\beta (\tilde{x})\tilde{z}}, C_{1} {=} \beta (\tilde{x})\tilde{z},\\
L_{1} {=} \tfrac{(\beta (\tilde{x}))^{3}\eta (\tilde{x},\tilde{z})\gamma_{1}(\tilde{x},\tilde{z})F_0(a,b)}{(\lambda_{1}(\tilde{x},\tilde{z}))^{2}}, L_{2} {=} \tfrac{\gamma_{1}(\tilde{x},\tilde{z})}{(d_{0}(\tilde{x},\tilde{z}))^{2}}.\\ \hline \end{array}$}
\label{fig:bcr6}
\end{figure}

\section{A minimal generating set for $\mathcal{Z}_{3}$}
\label{sec:mgsbc}

In Section \ref{sec:nmgs}, we presented generating sets for $\mathcal{Z}_{1,2}$ and $\mathcal{Z}_{2,1}$, the sets of impedances realized by series-parallel networks containing at most one capacitor and two inductors, and at most two capacitors and one inductor, respectively. However, these generating sets were not minimal. Moreover, the results of Section \ref{sec:nmgs} do not specify how to obtain a network realization for a given impedance from the classes $\mathcal{Z}_{1,2}$ or $\mathcal{Z}_{2,1}$. This motivates the two main contributions of this paper, which are presented in this section.

In our first notable contribution, we prove that the impedance of any series-parallel network containing at most three energy storage elements, with no constraint on the number of resistors, is also realized by a series-parallel network containing at most three energy storage elements and at most four resistors (Lemmas \ref{lem:z30mgs}--\ref{lem:z03mgs} and Theorems \ref{thm:z12mgs}--\ref{thm:z21mgs}). This represents a significant and non-trival extension to the famous theorem of Lin \cite{LIN}, which proved an analogous result for the simpler case of series-parallel networks containing at most two energy storage elements. It is also advantageous to the design of passive mechanical networks owing to the importance of minimizing their complexity, as will be discussed in greater detail in Section \ref{sec:pmc}. 

In our second notable contribution, we provide a systematic method for obtaining a minimal series-parallel network realization for any given impedance from the class $\mathcal{Z}_3$  (Theorems \ref{thm:mr3re}--\ref{thm:bqn14}). The significance of this result will be demonstrated in Section \ref{sec:pmc} using a practical example from \cite{fucheng_9b}. In general, in order to compute a realization for a given impedance from the set $\mathcal{Z}_3$ using the results of Theorems \ref{thm:mr3re}--\ref{thm:bqn14}, it is necessary to calculate the roots of a finite number of univariate polynomials, and to evaluate a number of other univariate polynomial functions at these roots. This is not computationally demanding; the example provided in this paper was solved in Maple 18 in under one minute on an HP EliteBook 840 G4 laptop.

\begin{table}[h]
\caption{Definition of network classes $\mathcal{N}_{16}$--$\mathcal{N}_{30}$. Refer to Figs.\ \ref{fig:bcr2}--\ref{fig:bcr6} for definitions of network classes $\mathcal{N}_{10}$--$\mathcal{N}_{14}$.}
\centering
\begin{tabular}{|p{2.5cm}|p{2.5cm}|p{2.5cm}|}
\hline
Network class& Originating \newline network class & Element removed \\
\hline
$\mathcal{N}_{16}$ & $\mathcal{N}_{10}$ & $R_{1}$ \\
\hline
$\mathcal{N}_{17}$ & $\mathcal{N}_{10}$ & $G_{3}$ \\
\hline
$\mathcal{N}_{18}$ & $\mathcal{N}_{10}$ & $R_{5}$ \\
\hline
$\mathcal{N}_{19}$ & $\mathcal{N}_{11}$ & $R_{1}$ \\
\hline
$\mathcal{N}_{20}$ & $\mathcal{N}_{11}$ & $G_{3}$ \\
\hline
$\mathcal{N}_{21}$ & $\mathcal{N}_{11}$ & $R_{5}$ \\
\hline
$\mathcal{N}_{22}$ & $\mathcal{N}_{12}$ & $G_{1}$ \\
\hline
$\mathcal{N}_{23}$ & $\mathcal{N}_{12}$ & $R_{3}$ \\
\hline
$\mathcal{N}_{24}$ & $\mathcal{N}_{12}$ & $R_{5}$ \\
\hline
$\mathcal{N}_{25}$ & $\mathcal{N}_{13}$ & $G_{1}$ \\
\hline
$\mathcal{N}_{26}$ & $\mathcal{N}_{13}$ & $R_{3}$ \\
\hline
$\mathcal{N}_{27}$ & $\mathcal{N}_{13}$ & $R_{5}$ \\
\hline
$\mathcal{N}_{28}$ & $\mathcal{N}_{14}$ & $R_{1}$ \\
\hline
$\mathcal{N}_{29}$ & $\mathcal{N}_{14}$ & $G_{3}$ \\
\hline
$\mathcal{N}_{30}$ & $\mathcal{N}_{14}$ & $R_{5}$ \\
\hline
\end{tabular}
\label{tab:n2047d}
\end{table}

Our minimal generating set for the class $\mathcal{Z}_3$ is summarized in the following two lemma and two theorem statements, which consider separately the classes (i) $\mathcal{Z}_{0,3}$, (ii) $\mathcal{Z}_{1,2}$, (iii) $\mathcal{Z}_{2,1}$, and (iv) $\mathcal{Z}_{3,0}$. Cases (i) and (iv) correspond to the well known Cauer canonical form and are included for completeness. Cases (ii) and (iii) are new to this paper, and will be proved in Section \ref{sec:proofs}.

\begin{lemma}
\label{lem:z30mgs}
The union of $\mathcal{N}_{1}$, $\mathcal{N}_{2}$, $\mathcal{N}_{4}$ and $\mathcal{N}_{15}$ is a minimal generating set for $\mathcal{Z}_{3,0}$ (see Figs.\ \ref{fig:c0r}, \ref{fig:bl1in}, \ref{fig:bq2co2i} and \ref{fig:bc3c}).
\end{lemma}

\begin{lemma}
\label{lem:z03mgs}
The union of $\mathcal{N}_{1}$, $\mathcal{N}_{3}$, $\mathcal{N}_{5}$ and $\mathcal{N}_{15}^{i}$ is a minimal generating set for $\mathcal{Z}_{0,3}$.
\end{lemma}

\begin{theorem}
\label{thm:z12mgs}
The union of $\mathcal{N}_{1}$--$\mathcal{N}_{3}$, $\mathcal{N}_{5}$--$\mathcal{N}_{9}$, $\mathcal{N}_{17}$, $\mathcal{N}_{18}$, $\mathcal{N}_{20}$--$\mathcal{N}_{27}$, $\mathcal{N}_{29}$, $\mathcal{N}_{30}$, $\mathcal{N}_{17}^{p}$, $\mathcal{N}_{18}^{p}$, $\mathcal{N}_{20}^{p}$--$\mathcal{N}_{27}^{p}$, $\mathcal{N}_{29}^{p}$ and $\mathcal{N}_{30}^{p}$ is a minimal generating set for $\mathcal{Z}_{1,2}$ (see Figs.\ \ref{fig:c0r}--\ref{fig:bq1c1l} and Table \ref{tab:n2047d}).
\end{theorem}

\begin{theorem}
\label{thm:z21mgs}
The union of $\mathcal{N}_{1}$--$\mathcal{N}_{4}$, $\mathcal{N}_{6}$--$\mathcal{N}_{9}$, $\mathcal{N}_{17}^{d}$, $\mathcal{N}_{18}^{d}$, $\mathcal{N}_{20}^{d}$--$\mathcal{N}_{27}^{d}$, $\mathcal{N}_{29}^{d}$, $\mathcal{N}_{30}^{d}$, $\mathcal{N}_{17}^{i}$, $\mathcal{N}_{18}^{i}$, $\mathcal{N}_{20}^{i}$--$\mathcal{N}_{27}^{i}$, $\mathcal{N}_{29}^{i}$ and $\mathcal{N}_{30}^{i}$ is a minimal generating set for $\mathcal{Z}_{2,1}$.
\end{theorem}

Lemmas \ref{lem:z30mgs} and \ref{lem:z03mgs} are well known, and we refer to \cite{HugCF} for a proof.
Indeed, explicit realizations for any impedance from one of the classes $\mathcal{Z}_{3,0}$ or $\mathcal{Z}_{0,3}$ were provided in \cite{HugCF}, and are summarized in the following lemma for completeness.
\begin{lemma}
\label{lem:3resc}
Let $Z \in \mathcal{Z}_{3,0}$. If $F_0(a,b) \neq 0$ (i.e., $Z$ is bicubic), then $a_3, b_0 \geq 0$ and $\abs{\mathcal{S}_i(a,b)} > 0$ for $i = 1, \ldots , 6$, and $Z$ is the impedance of the network $N_{15}$ in Fig.\ \ref{fig:bc3c} when the element values are as indicated in that figure's caption.

If, on the other hand, $F_0(a,b) = 0$, then $Z \in \mathcal{Z}_{2,0}$.
\end{lemma}

\tikzset{ ccbc1/.pic={
\ctikzset{bipoles/thickness=1}
\node[anchor=center] at (1,0) (O) {};
\draw (O) to[short,*-] ++(0.3,0)
to[R=$R_{1}$] ++(1.7,0)
to[short] ++(0,0.7)
to[R=$R_{2}$] ++(1.7,0)
to[short] ++(0,0.7)
to[R=$R_{3}$] ++(1.7,0)
to[short] ++(0,0.7)
to[R=$\frac{1}{G_{4}}$] ++(1.7,0)
to[short] ++(0,-2.1)
($(O)+(3.7,0.7)$) to[short] ++(0,-0.7)
to[C=$\frac{1}{C_{2}s}$] ++(1.7,0)
to[short] ++(1.7,0)
to[short,-*] ++(0.3,0)
($(O)+(5.4,1.4)$) to[short] ++(0,-0.7)
to[C=$\frac{1}{C_{1}s}$] ++(1.7,0)
($(O)+(2.0,0)$) to[short] ++(0,-0.7)
to[C=$\frac{1}{C_{3}s}$] ++(1.7,0)
to[short] ++(3.4,0)
to[short] ++(0,0.7);}}

\tikzset{ llbc1/.pic={
\ctikzset{bipoles/thickness=1}
\node[anchor=center] at (1,0) (O) {};
\draw (O) to[short,*-] ++(0.3,0)
to[R=$R_{1}$] ++(1.7,0)
to[short] ++(0,0.7)
to[R=$R_{2}$] ++(1.7,0)
to[short] ++(0,0.7)
to[R=$R_{3}$] ++(1.7,0)
to[short] ++(0,0.7)
to[R=$\frac{1}{G_{4}}$] ++(1.7,0)
to[short] ++(0,-2.1)
($(O)+(3.7,0.7)$) to[short] ++(0,-0.7)
to[L=$L_{2}$] ++(1.7,0)
to[short] ++(1.7,0)
to[short,-*] ++(0.3,0)
($(O)+(5.4,1.4)$) to[short] ++(0,-0.7)
to[L=$L_{1}s$] ++(1.7,0)
($(O)+(2.0,0)$) to[short] ++(0,-0.7)
to[L=$L_{3}s$] ++(1.7,0)
to[short] ++(3.4,0)
to[short] ++(0,0.7);}}

\begin{figure}[!b]
\centering
\normalsize
\begin{tikzpicture}[scale=0.57, every node/.style={transform shape}, font=\Large]
\node at (-1.6,2.5) {$N_{15}$};
\node at (-1.6,0.0) {$\begin{array}{rl}\text{(i)}& R_{1} \geq 0\\
\text{(ii)}& R_{2}, R_{3} > 0\\
\text{(iii)}& G_{4} \geq 0\\
\text{(iv)}& C_{1}, C_{2}, C_{3} > 0
\end{array}$};
\pic at (0,0) [] {ccbc1};
\end{tikzpicture}
\caption[]{Network $N_{15}$. We define $\mathcal{N}_{15}$ as the set of all networks of the form of $N_{15}$ that satisfy conditions (i)--(iv). \\[0.2em] 

\quad If the conditions of Lemma \ref{lem:3resc} hold, then $N_{15}$ has impedance $Z = a/b$ with $a, b \in \mathbb{R}[s]$ as in (\ref{eq:abd}) when the element values are:\\[0.2cm] $\begin{array}{|l|}\hline \rule{0pt}{1.1\normalbaselineskip}R_{1} = \tfrac{a_3}{\abs{\mathcal{S}_1(a,b)}}, R_{2} = \tfrac{\abs{\mathcal{S}_2(a,b)}^{2}}{\abs{\mathcal{S}_1(a,b)}\abs{\mathcal{S}_3(a,b)}}, R_{3} = \tfrac{\abs{\mathcal{S}_4(a,b)}^2}{\abs{\mathcal{S}_3(a,b)}\abs{\mathcal{S}_5(a,b)}}, \\
G_{4} = \tfrac{\abs{\mathcal{S}_5(a,b)}b_0}{\abs{\mathcal{S}_6(a,b)}}, C_{1} = \tfrac{\abs{\mathcal{S}_5(a,b)}^2}{\abs{\mathcal{S}_4(a,b)}\abs{\mathcal{S}_6(a,b)}}, C_{2} = \tfrac{\abs{\mathcal{S}_3(a,b)}^2}{\abs{\mathcal{S}_2(a,b)}\abs{\mathcal{S}_4(a,b)}}, \\
C_{3} = \tfrac{\abs{\mathcal{S}_1(a,b)}^2}{\abs{\mathcal{S}_2(a,b)}} \\ \hline \end{array}$}.
\label{fig:bc3c}
\end{figure}

In this paper, we will prove Theorem \ref{thm:z12mgs}, and Theorem \ref{thm:z21mgs} can be proved in a similar manner. We first show that the impedance of any network from the generating set described in Theorem \ref{thm:z12nmgs} can also be realized by a network from one of the classes $\mathcal{N}_1$--$\mathcal{N}_3$, $\mathcal{N}_5$--$\mathcal{N}_9$, $\mathcal{N}_{16}$--$\mathcal{N}_{30}$ or $\mathcal{N}_{16}^p$--$\mathcal{N}_{30}^p$ (see Table \ref{tab:n2047d}). Finally, we show that the impedance of any network from $\mathcal{N}_{16}$, $\mathcal{N}_{19}$, $\mathcal{N}_{28}$, $\mathcal{N}_{16}^{p}$, $\mathcal{N}_{19}^{p}$ or $\mathcal{N}_{28}^p$ can also be realized by a network from one of the other network classes listed in Theorem \ref{thm:z12mgs}. In the course of the proof, we obtain explicit realizations for any impedance from the class $\mathcal{Z}_{1,2}$. Together with Lemma \ref{lem:3resc} and the results of Section \ref{sec:ncm}, we thus obtain an explicit realization for any given  $Z \in \mathcal{Z}_3$, in accordance with the following theorem.

\begin{theorem}
\label{thm:mr3re}
If $Z \in \mathcal{Z}_3$, then $Z$ is the impedance of a network from one of the classes (i) $\mathcal{N}_1$--$\mathcal{N}_3$; (ii) $\mathcal{N}_4$--$\mathcal{N}_9$; (iii) $\mathcal{N}_{10}$--$\mathcal{N}_{14}$; (iv) $\mathcal{N}_{10}^i$--$\mathcal{N}_{14}^i$; (v) $\mathcal{N}_{10}^p$--$\mathcal{N}_{14}^p$; (vi) $\mathcal{N}_{10}^d$--$\mathcal{N}_{14}^d$; (vii) $\mathcal{N}_{15}$; or (viii) $\mathcal{N}_{15}^i$.

Corresponding to each of the above cases, a series-parallel generic network $N$ realizing $Z$ is obtained as follows:
\begin{enumerate}[label=(\roman*)]
\item $N$ is obtained from Lemma \ref{lem:blgs};
\item $N$ is obtained from Lemma \ref{lem:bqrrc};
\item $N = \hat{N}$ where $\hat{N}$ is obtained from either (i) Theorem \ref{thm:bcn1013} or \ref{thm:bcn14} (if $Z$ is bicubic); (ii) Lemma \ref{lem:bqrrc} or Theorem \ref{thm:bqn1013} or \ref{thm:bqn14} (if $Z$ is biquadratic); or (iii) Lemma \ref{lem:blgs} (if $Z$ is bilinear or constant);\label{nl:frcc5}
\item $N = \hat{N}^i$ where $\hat{N}$ is obtained as in condition \ref{nl:frcc5};
\item $N = \hat{N}^p$ where $\hat{N}$ is obtained as in condition \ref{nl:frcc5};
\item $N = \hat{N}^d$ where $\hat{N}$ is obtained as in condition \ref{nl:frcc5};
\item $N$ is obtained from Lemma \ref{lem:3resc};
\item $N = \hat{N}^i$ where $\hat{N}$ is obtained from Lemma \ref{lem:3resc}.
\end{enumerate}
\end{theorem}

Our proof considers separately the cases in which the impedance is (a) bicubic, (b) biquadratic, and (c) bilinear or constant. In case (a), an explicit realization for any impedance realized by a network from one of the classes $\mathcal{N}_{10}$--$\mathcal{N}_{13}$ (resp., $\mathcal{N}_{14}$) is provided in Theorem \ref{thm:bcn1013} (resp., Theorem \ref{thm:bcn14}) in terms of the polynomials
\begin{align}
a(s) &= a_{3}s^{3} + a_{2}s^{2} + a_{1}s + a_{0}, \text{ and} \nonumber \\
b(s) &= b_{3}s^{3} + b_{2}s^{2} + b_{1}s + b_{0}, \label{eq:abd}
\end{align}
in a coprime factorization $Z = a/b$ for the impedance $Z$. Also, Theorems \ref{thm:bqn1013} and \ref{thm:bqn14} provide explicit realizations in case (b) in terms of the polynomials
\begin{align}
a(s) &= a_{2}s^{2} + a_{1}s + a_{0}, \text{ and} \nonumber \\
b(s) &= b_{2}s^{2} + b_{1}s + b_{0},\label{eq:abd2}
\end{align}
in a coprime factorization $Z = a/b$ of the impedance $Z$. Finally, in case (c), since the impedance $Z$ is necessarily positive-real, then it follows from Lemmas \ref{lem:blgs}--\ref{lem:blec} that $Z$ is also realized by a network from one of the classes $\mathcal{N}_1$--$\mathcal{N}_3$.

\begin{theorem}
\label{thm:bcn1013}
Let $a,b$ in (\ref{eq:abd}) be coprime; let $b_{i} > 0$ for at least one value of $i \in 0, \ldots , 3$; let $Z = a/b$; let $\alpha $, $\beta $, $c_0$--$c_2$, $d_0$--$d_2$ and $\kappa \in \mathbb{R}[u]$ be as in row (A2) of Table \ref{tab:term}; and let $\gamma_1$--$\gamma_3$ and $\lambda_1$--$\lambda_4 \in \mathbb{R}[u]$, be as in row (A1) of that table. If $Z$ is the impedance of a network from one of the classes $\mathcal{N}_{10}$--$\mathcal{N}_{13}$, then there exists an $x \geq 0$ such that one of the following pairs of conditions hold (see Table \ref{tab:cons} in the appendix): 
\begin{enumerate}[label=(\alph*)]
\item (C1) and (E1); \label{nl:bcrev1}
\item (C2) and (E2); \label{nl:bcrev2}
\item (C3) and (E3); or \label{nl:bcrev3}
\item (C4) and (E4). \label{nl:bcrev4}
\end{enumerate}
In each case, an explicit realization for $Z$ is obtained as follows:
\begin{enumerate}
\item If $\tilde{x} \geq 0$ is such that condition (C1) holds, then $\beta (\tilde{x}), d_{0}(\tilde{x}), d_{2}(\tilde{x}) \neq 0$, $\kappa (\tilde{x}), \gamma_{1}(\tilde{x}) > 0$, and $Z$ is the impedance of $N_{10}$ in Fig.\ \ref{fig:bcr2} when the element values are as indicated in the uppermost table of that figure's caption. In particular, if condition (E1)(i) (resp., (E1)(ii), (E1)(iii)) holds, then $R_5 = 0$ (resp., $G_3 = 0$, $R_1 = 0$), so $N_{10}$ is a network from class $\mathcal{N}_{18}$ (resp., $\mathcal{N}_{17}$, $\mathcal{N}_{16}$).\label{nl:bcev1}
\item If $\tilde{x} \geq 0$ is such that condition (C2) holds, then $\beta (\tilde{x}), d_{0}(\tilde{x}), d_{2}(\tilde{x}) \neq 0$, $\kappa (\tilde{x}), -\gamma_{3}(\tilde{x}) > 0$, and $Z$ is the impedance of $N_{11}$ in Fig.\ \ref{fig:bcr2} when the element values are as indicated in the uppermost table of that figure's caption. In particular, if condition (E2)(i) (resp., (E2)(ii), (E2)(iii)) holds, then $R_5 = 0$ (resp., $G_3 = 0$, $R_1 = 0$), so $N_{11}$ is a network from class $\mathcal{N}_{21}$ (resp., $\mathcal{N}_{20}$, $\mathcal{N}_{19}$).\label{nl:bcev2}
\item If $\tilde{x} \geq 0$ is such that condition (C3) holds, then $\beta (\tilde{x}), c_{0}(\tilde{x}), c_{2}(\tilde{x}) \neq 0$, $\kappa (\tilde{x}), -\gamma_{1}(\tilde{x}) > 0$, and $Z$ is the impedance of $N_{12}$ in Fig.\ \ref{fig:bcr5} when the element values are as indicated in the uppermost table of that figure's caption. In particular, if condition (E3)(i) (resp., (E3)(ii), (E3)(iii)) holds, then $R_5 = 0$ (resp., $R_3 = 0$, $G_1 = 0$), so $N_{12}$ is a network from class $\mathcal{N}_{24}$ (resp., $\mathcal{N}_{23}$, $\mathcal{N}_{22}$).\label{nl:bcev3}
\item If $\tilde{x} \geq 0$ is such that condition (C4) holds, then $\beta (\tilde{x}), c_{0}(\tilde{x}), c_{2}(\tilde{x}) \neq 0$, $\kappa (\tilde{x}), \gamma_{3}(\tilde{x}) > 0$, and $Z$ is the impedance of $N_{13}$ in Fig.\ \ref{fig:bcr5} when the element values are as indicated in the uppermost table of that figure's caption. In particular, if condition (E4)(i) (resp., (E4)(ii), (E4)(iii)) holds, then $R_5 = 0$ (resp., $R_3 = 0$, $G_1 = 0$), so $N_{13}$ is a network from class $\mathcal{N}_{27}$ (resp., $\mathcal{N}_{26}$, $\mathcal{N}_{25}$).\label{nl:bcev4}
\end{enumerate}
\end{theorem}

\begin{theorem}
\label{thm:bcn14}
Let $a,b$ in (\ref{eq:abd}) be coprime; let $b_{i} > 0$ for at least one value of $i \in 0, \ldots , 3$; let $Z = a/b$; let $\alpha $, $\beta $, $c_0$--$c_2$, $d_0$--$d_2$ and $\kappa \in \mathbb{R}[u]$ be as in row (A3) of Table \ref{tab:term}; and let $\gamma_1$--$\gamma_3$ and $\lambda_1$--$\lambda_4 \in \mathbb{R}[u]$ be as in row (A1) of that table. If $Z$ is the impedance of a network from $\mathcal{N}_{14}$, then there exists an $\tilde{x} \geq 0$ such that conditions (C5) and (E5) hold (see Table \ref{tab:cons}). 

In this case, an explicit realization for $Z$ is obtained as follows. If $\tilde{x} \geq 0$ is such that (C5) holds, then $\beta (\tilde{x}), d_{0}(\tilde{x}), \lambda_{1}(\tilde{x}) \neq 0$, $\kappa (\tilde{x}), \gamma_{1}(\tilde{x}) > 0$, and $Z$ is the impedance of $N_{14}$ in Fig.\ \ref{fig:bcr6} when the element values are as indicated in the uppermost table of that figure's caption. In particular, if condition (E5)(i) (resp., (E5)(ii), (E5)(iii)) holds, then $R_5 = 0$ (resp., $R_1 = 0$, $G_3 = 0$), so $N_{14}$ is a network from class $\mathcal{N}_{30}$ (resp., $\mathcal{N}_{28}$, $\mathcal{N}_{29}$).
\end{theorem}

\begin{theorem}
\label{thm:bqn1013}
Let $a,b$ in (\ref{eq:abd2}) be coprime; let $b_{i} > 0$ for at least one value of $i \in 0, 1, 2$; let $Z = a/b$; let $\alpha $, $\beta $, $c_0$--$c_2 \in \mathbb{R}[u]$, $d_0$--$d_2$ and $\eta \in \mathbb{R}[u,v]$ be as in row (A4) of Table \ref{tab:term}; and let $\gamma_1$--$\gamma_3$ and $\lambda_1$--$\lambda_4 \in \mathbb{R}[u,v]$ be as in row (A1) of that table. If $Z$ is the impedance of a network from one of the classes $\mathcal{N}_{10}$--$\mathcal{N}_{13}$, then either $Z \in \mathcal{Z}_{1,1}$, $Z \in \mathcal{Z}_{0,2}$, or there exist $\tilde{x} \geq 0$ and $\tilde{z} \neq 0$ such that one of the following pairs of conditions hold (see Table \ref{tab:cons}): 
\begin{enumerate}[label=(\alph*)]
\item (Q5) and (F1); 
\item (Q6) and (F2); 
\item (Q7) and (F3); or 
\item (Q8) and (F4).
\end{enumerate}
If $Z \in \mathcal{Z}_{1,1}$ or $Z \in \mathcal{Z}_{0,2}$, an explicit realization for $Z$ is obtained from Lemma \ref{lem:bqrrc}. Otherwise, an explicit realization for $Z$ is obtained as follows:
\begin{enumerate}[leftmargin=0.3cm]
\item If $\tilde{x} \geq 0$ and $\tilde{z} \neq 0$ are such that (Q5) holds, then $\beta (\tilde{x}), d_{0}(\tilde{x},\tilde{z}), d_{2}(\tilde{x},\tilde{z}) \neq 0$, $\gamma_{1}(\tilde{x},\tilde{z}) > 0$, and $Z$ is the impedance of $N_{10}$ in Fig.\ \ref{fig:bcr2} when the element values are as indicated in the lowermost table of that figure's caption. In particular, if condition (F1) holds, then $G_3 = 0$, so $N_{10}$ is a network from class $\mathcal{N}_{17}$.
\item If $\tilde{x} \geq 0$ and $\tilde{z} \neq 0$ are such that (Q6) holds, then $\beta (\tilde{x}), d_{0}(\tilde{x},\tilde{z}), d_{2}(\tilde{x},\tilde{z}) \neq 0$, $\gamma_{3}(\tilde{x},\tilde{z}) < 0$, and $Z$ is the impedance of $N_{11}$ in Fig.\ \ref{fig:bcr2} when the element values are as indicated in the lowermost table of that figure's caption. In particular, if condition (F2) holds, then $G_3 = 0$, so $N_{11}$ is a network from class $\mathcal{N}_{20}$.
\item If $\tilde{x} \geq 0$ and $\tilde{z} \neq 0$ are such that (Q7) holds, then $\beta (\tilde{x}), c_{0}(\tilde{x}), c_{2}(\tilde{x}) \neq 0$, $\gamma_{1}(\tilde{x},\tilde{z}) < 0$, and $Z$ is the impedance of $N_{12}$ in Fig.\ \ref{fig:bcr5} when the element values are as indicated in the lowermost table of that figure's caption. In particular, if condition (F3)(i) (resp., (F3)(ii)) holds, then $R_3 = 0$ (resp., $G_1 = 0$), so $N_{12}$ is a network from class $\mathcal{N}_{23}$ (resp., $\mathcal{N}_{22}$).
\item If $\tilde{x} \geq 0$ and $\tilde{z} \neq 0$ are such that (Q8) holds, then $\beta (\tilde{x}), c_{0}(\tilde{x}), c_{2}(\tilde{x}) \neq 0$, $\gamma_{3}(\tilde{x},\tilde{z}) > 0$, and $Z$ is the impedance of $N_{13}$ in Fig.\ \ref{fig:bcr5} when the element values are as indicated in the lowermost table of that figure's caption. In particular, if condition (F4)(i) (resp., (F4)(ii)) holds, then $R_3 = 0$ (resp., $G_1 = 0$), so $N_{13}$ is a network from class $\mathcal{N}_{26}$ (resp., $\mathcal{N}_{25}$).
\end{enumerate}
\end{theorem}

\begin{theorem}
\label{thm:bqn14}
Let $a,b$ in (\ref{eq:abd2}) be coprime; let $b_{i} > 0$ for at least one value of $i \in 0, 1, 2$; let $Z = a/b$; let $\alpha $, $\beta $, $c_0$--$c_2 \in \mathbb{R}[u]$ $d_0$--$d_2$ and $\eta \in \mathbb{R}[u,v]$ be as in row (A5) of Table \ref{tab:term}; and let $\gamma_1$--$\gamma_3$ and $\lambda_1$--$\lambda_4 \in \mathbb{R}[u,v]$ be as in row (A1) of that table. If $Z$ is the impedance of a network from $\mathcal{N}_{14}$, then either $Z \in \mathcal{Z}_{1,1}$, $Z \in \mathcal{Z}_{0,2}$, or there exist $\tilde{x} \geq 0$ and $\tilde{z} \neq 0$ such that conditions (Q11) and (F5) hold (see Table \ref{tab:cons}). 

In this case, an explicit realization for $Z$ is obtained as follows. If $\tilde{x} \geq 0$ is such that (Q11) holds, then $\beta (\tilde{x}), d_{0}(\tilde{x},\tilde{z}), \lambda_{1}(\tilde{x},\tilde{z}) \neq 0$, $\gamma_{1}(\tilde{x},\tilde{z}) > 0$, and $Z$ is the impedance of $N_{14}$ in Fig.\ \ref{fig:bcr6} when the element values are as indicated in the lowermost table of that figure's caption. In particular, if condition (F5) holds, then $G_3 = 0$, so $N_{14}$ is a network from class $\mathcal{N}_{29}$.
\end{theorem}

\begin{remark}
\label{rem:ercc}
In \cite{ZJ_ERBI}, the concept of an \emph{essential-regular} function was defined, and a minimal generating set was obtained for the class of bicubic essential-regular functions. However, not every impedance $Z \in \mathcal{Z}_{3}$ is essential-regular. To see this, note initially from \cite{ZJ_ERBI} that if $Z = a/b$ where $a, b \in \mathbb{R}[s]$ are coprime polynomials as in (\ref{eq:abd}), and if $Z$ is essential regular, then either (i) $a_{3}b_{2} - a_{2}b_{3}$, $a_{3}b_{1} - a_{1}b_{3}$ and $a_{3}b_{0} - a_{0}b_{3}$ have the same sign; or (ii) $a_{3}b_{0} - a_{0}b_{3}$, $a_{2}b_{0} - a_{0}b_{2}$ and $a_{1}b_{0} - a_{0}b_{1}$ have the same sign. Now, consider the impedance of network $N_{10}$ in Fig.\ \ref{fig:bcr2}, and let $R_{1} = \tfrac{4}{9}$, $R_{2} = \tfrac{128}{279}$, $G_{3} = \tfrac{31}{96}$, $R_{4} = 0$, $R_{5} = 1$, $C_{1} = \tfrac{405}{64}$, $L_{1} = \tfrac{576}{961}$, and $L_{2} = 1$. It can be verified that the impedance coefficients for this network are $a_{3} = 60$, $a_{2} = 124$, $a_{1} = 92$, $a_{0} = 28$, $b_{3} = 135$, $b_{2} = 291$, $b_{1} = 191$ and $b_{0} = 59$, and by direct calculation we find that neither (i) nor (ii) hold, so $Z$ is not essential regular.
\end{remark}

\section{A minimal generating set for $\mathcal{Z}_3$: proofs}
\label{sec:proofs}

This section contains the proofs of Theorems \ref{thm:z12mgs}, \ref{thm:z21mgs} and \ref{thm:mr3re}--\ref{thm:bqn14}. We first prove Theorem \ref{thm:bcn1013}. The proofs of Theorems \ref{thm:bcn14}--\ref{thm:bqn14} are similar and are omitted for brevity. Then, we prove Theorem \ref{thm:mr3re}. Finally, we prove Theorem \ref{thm:z12mgs}, and Theorem \ref{thm:z21mgs} can be proved in an analogous manner.

In order to prove Theorem \ref{thm:bcn1013}, it will be shown that $Z$ is the impedance of a network from one of the classes $\mathcal{N}_{10}$--$\mathcal{N}_{13}$ if and only if there exists an $\tilde{x} \geq 0$ such that one of the conditions (C1)--(C4) in that theorem hold, and $\tilde{x}$ can be varied continuously such that at least one of these conditions continues to hold up to the point that one of the pairs of conditions \ref{nl:bcrev1}--\ref{nl:bcrev4} holds. This then results in a network realization for $Z$ from one of the network classes $\mathcal{N}_{16}$--$\mathcal{N}_{27}$. The proof is facilitated by the next two lemma that characterise the impedances realized by networks from classes $\mathcal{N}_{10}$--$\mathcal{N}_{13}$ in terms of conditions (C1)--(C4) in Theorem \ref{thm:bcn1013}, and establish some useful equivalent conditions.
\begin{lemma}
\label{lem:bcrc1}
Let $a,b$ in (\ref{eq:abd}) be coprime; let $b_{i} > 0$ for at least one value of $i \in 0, \ldots , 3$; let $Z = a/b$; let $\alpha $, $\beta $, $c_0$--$c_2$, $d_0$--$d_2$ and $\kappa \in \mathbb{R}[u]$ be as in row (A2) of Table \ref{tab:term}; and let $\gamma_1$--$\gamma_3$ and $\lambda_1$--$\lambda_4 \in \mathbb{R}[u]$ be as in row (A1) of that table. If $Z$ is the impedance of $N_{10}$ (resp., $N_{11}$, $N_{12}$, $N_{13}$), then there exists $\tilde{x} \geq 0$ such that the element values take the form indicated in condition \ref{nl:bcev1} (resp., \ref{nl:bcev2}, \ref{nl:bcev3}, \ref{nl:bcev4}) of Theorem \ref{thm:bcn1013}.
\end{lemma}

\begin{IEEEproof}
We first consider the case where $Z$ is the impedance of $N_{10}$. With $\tilde{x} = R_{5}/L_{2}$, then $\tilde{x} \geq 0$, and we note that
\begin{align}
Z(s) &= R_{4} + \tfrac{1}{\frac{1}{Z_{2}(s)} + \frac{1}{L_{2}(s+\tilde{x})}} \nonumber \\
&= R_{4} + \frac{Z_{2}(s)L_{1}(s+\tilde{x})}{Z_{2}(s) + L_{2}(s+\tilde{x})},\label{eq:zbcgf}
\end{align}
where $Z_{2}$ is the impedance of a network of the form of $N_{6}$ in Fig.\ \ref{fig:bq1c1l}. Since $Z$ is bicubic, then $Z_{2}$ must be biquadratic and the poles of $1/Z_{2}(s)$ and $1/(L_{2}(s+\tilde{x}))$ must be distinct, which implies that $Z_{2}(-\tilde{x}) \neq 0$. Then, from (\ref{eq:zbcgf}), it follows that $R_{4} = Z(-\tilde{x})$, and so $R_4$ has the form indicated in item \ref{nl:bcev1} of Theorem \ref{thm:bcn1013}. Also, with the notation $c(s) = c_2(\tilde{x})s^2 + c_1(\tilde{x})s + c_0(\tilde{x})$ and $d(s) = d_2(\tilde{x}) s^2 + d_1(\tilde{x}) s + d_0(\tilde{x})$, then we note that $c(s) = \begin{bmatrix}1& -\tilde{x}& \tilde{x}^2\end{bmatrix}\mathcal{B}(a,b)\begin{bmatrix}1& s& s^2\end{bmatrix}^T$ and $d(s) = \beta(-\tilde{x})\begin{bmatrix}1& -\tilde{x}& \tilde{x}^2\end{bmatrix}\mathcal{B}(b,c)\begin{bmatrix}1& s& s^2\end{bmatrix}^T$, and it follows from Definition \ref{def:bezd} that
\begin{align}
(s+\tilde{x})c(s) &= a(s)b(-\tilde{x})-b(s)a(-\tilde{x}), \text{ and}\label{eq:cbezr}\\
(s+\tilde{x})d(s) &= b(-\tilde{x})(b(s)c(-\tilde{x})-b(-\tilde{x})c(s)),\label{eq:dbezr}
\end{align}
whereupon we conclude that
\begin{align}
\frac{1}{Z_{2}(s)} &= \frac{1}{Z(s)-Z(-\tilde{x})} - \frac{1}{L_{2}(s+\tilde{x})} \\
&= \frac{b(s)b(-\tilde{x})}{(s+\tilde{x})c(s)} - \frac{1}{L_{2}(s+\tilde{x})}.\label{eq:zbqrgf}
\end{align}
By multiplying both sides in the above equation by $s+\tilde{x}$ and taking the limit as $s \rightarrow -\tilde{x}$, we find that $L_{2}$ and $R_{5}$ also have the forms indicated in item \ref{nl:bcev1} of Theorem \ref{thm:bcn1013}. Moreover, (\ref{eq:zbqrgf}) implies that
\begin{equation}
\hspace*{-0.4cm} \frac{1}{Z_{2}(s)} = \frac{b(-\tilde{x})(b(s)c(-\tilde{x})-c(s)b(-\tilde{x}))}{(s+\tilde{x})c(s)c(-\tilde{x})} = \frac{d(s)}{c(s)c(-\tilde{x})}.
\end{equation}
Thus, with the notation $f(s) = c(-\tilde{x})c(s) = \kappa (\tilde{x})c(s)$ and $g(s) =d(s)$, we find that $Z_{2} = f/g$. Direct calculation then gives $F_0(f,g) = -(\beta (\tilde{x}))^{4}(\kappa (\tilde{x}))^{3}F_0(a,b)$. Since, in addition, $Z_{2}$ is the impedance of a network of the form of $N_{6}$, then from Lemma \ref{lem:bqrrc} and Remark \ref{rem:z2lo2c}, we conclude that $R_{1}, R_{2}, G_{3}, C_{1}$ and $L_{1}$ have the forms indicated in item \ref{nl:bcev1} of Theorem \ref{thm:bcn1013}.

The case in which $Z$ is the impedance of $N_{11}$ (resp., $N_{12}$, $N_{13}$) is similar. In this case, $Z_{2}$ is the impedance of a network of the form of $N_{7}$ (resp., $N_{8}, N_{9}$). 
\end{IEEEproof}

\begin{lemma}
\label{lem:bciec}
Let $a,b$ in (\ref{eq:abd}) be coprime; let $b_{i} > 0$ for at least one value of $i \in 0, \ldots , 3$; let $\alpha $, $\beta $, $c_0$--$c_2$, $d_0$--$d_2$ and $\kappa \in \mathbb{R}[u]$ be as in row (A2) of Table \ref{tab:term}; let $\gamma_1$--$\gamma_3$ and $\lambda_1$--$\lambda_4 \in \mathbb{R}[u]$ be as in row (A1) of that table; let (C1)--(C4) be the sets of inequalities in Table \ref{tab:cons}; and consider the following additional sets of inequalities:
\begin{remunerate}
\labitem{(C\Alph{muni})}{nl:bcca1} $b_{i} > 0$ for at least one value of $i \in 0,\ldots,3$; $\alpha (\tilde{x})$ and $\beta (\tilde{x})$ have the same sign; $\beta (\tilde{x}) \neq 0$; $c_{0}(\tilde{x})d_{0}(\tilde{x})$, $d_{2}(\tilde{x})\lambda_{1}(\tilde{x})$, $\gamma_{2}(\tilde{x})d_{0}(\tilde{x})d_{2}(\tilde{x}) \geq 0$; $d_{0}(\tilde{x}), d_{2}(\tilde{x}) \neq 0$; $\kappa (\tilde{x}), \gamma_{1}(\tilde{x}) > 0$; and $F_0(a,b) > 0$.
\labitem{(C\Alph{muni})}{nl:bcca2} $b_{i} > 0$ for at least one value of $i \in 0,\ldots,3$; $\alpha (\tilde{x})$ and $\beta (\tilde{x})$ have the same sign; $\beta (\tilde{x}) \neq 0$; $c_{2}(\tilde{x})d_{2}(\tilde{x})$, $d_{0}(\tilde{x})\lambda_{3}(\tilde{x})$, $-\gamma_{2}(\tilde{x})d_{0}(\tilde{x})d_{2}(\tilde{x}) \geq 0$; $d_{0}(\tilde{x}), d_{2}(\tilde{x}) \neq 0$; $\kappa (\tilde{x}), -\gamma_{3}(\tilde{x}) < 0$; and $F_0(a,b) > 0$.
\labitem{(C\Alph{muni})}{nl:bcca3} $b_{i} > 0$ for at least one value of $i \in 0,\ldots,3$; $\alpha (\tilde{x})$ and $\beta (\tilde{x})$ have the same sign; $\beta (\tilde{x}) \neq 0$; $c_{0}(\tilde{x})d_{0}(\tilde{x})$, $c_{2}(\tilde{x})\lambda_{4}(\tilde{x})$, $-\gamma_{2}(\tilde{x})c_{0}(\tilde{x})c_{2}(\tilde{x}) \geq 0$; $c_{0}(\tilde{x}), c_{2}(\tilde{x}) \neq 0$; $\kappa (\tilde{x}), -\gamma_{1}(\tilde{x}) > 0$; and $F_0(a,b) > 0$.
\labitem{(C\Alph{muni})}{nl:bcca4} $b_{i} > 0$ for at least one value of $i \in 0,\ldots,3$; $\alpha (\tilde{x})$ and $\beta (\tilde{x})$ have the same sign; $\beta (\tilde{x}) \neq 0$; $c_{2}(\tilde{x})d_{2}(\tilde{x})$, $c_{0}(\tilde{x})\lambda_{2}(\tilde{x})$, $\gamma_{2}(\tilde{x})c_{0}(\tilde{x})c_{2}(\tilde{x}) \geq 0$; $c_{0}(\tilde{x}), c_{2}(\tilde{x}) \neq 0$; $\kappa (\tilde{x}), \gamma_{3}(\tilde{x}) > 0$; and $F_0(a,b) > 0$.
\end{remunerate}
Then (C1) (resp., (C2), (C3), (C4)) is satisfied if and only if \ref{nl:bcca1} (resp., \ref{nl:bcca2}, \ref{nl:bcca3}, \ref{nl:bcca4}) is satisfied.
\end{lemma}

\begin{IEEEproof}
We first let condition (C1) hold and we prove that so too does condition \ref{nl:bcca1}. Since $b_{i} \geq 0$ for $i = 0,1,2,3$ and $F_0(a,b) \neq 0$, then $b_{i} > 0$ for at least one value of $i \in 0, \ldots , 3$. Next, we let $c(s) = c_2(\tilde{x}) s^2 + c_1(\tilde{x}) s + c_0(\tilde{x})$ and $d(s) = d_2(\tilde{x}) s^2 + d_1(\tilde{x}) s + d_0(\tilde{x})$, we recall the relationships (\ref{eq:cbezr}) and (\ref{eq:dbezr}), and we note that $\beta(\tilde{x}) = b(-\tilde{x})$ and $\alpha(\tilde{x}) = a(-\tilde{x})$. From (\ref{eq:cbezr}), if $\beta (\tilde{x}) = 0$, then $c(s)(s+\tilde{x}) = -b(s)\alpha (\tilde{x})$. Since $c_{0}(\tilde{x}), c_{1}(\tilde{x}), c_{2}(\tilde{x})$ and $\alpha (\tilde{x})$ have the same sign, $\tilde{x} \geq 0$, and $b_{i} \geq 0$ for $i = 0,1,2,3$, then we require $c_{0}(\tilde{x}) = c_{1}(\tilde{x}) = c_{2}(\tilde{x}) = 0$, so $\begin{bmatrix}1& -\tilde{x}& \tilde{x}^{2}\end{bmatrix}\mathcal{B}(a, b) = 0$. This implies that $\abs{\mathcal{B}(a, b)} = F_0(b,a) = 0$, whereupon $F_0(a,b) = 0$. But $F_0(a,b) > 0$, so we conclude that $\beta (\tilde{x}) \neq 0$. Next, from (\ref{eq:dbezr}), it follows that if $\kappa (\tilde{x}) = 0$ then $d(s)(s+\tilde{x}) = -\beta(\tilde{x})^{2}c(s)$. Since $c_{0}(\tilde{x}), c_{1}(\tilde{x}), c_{2}(\tilde{x}), d_{0}(\tilde{x}), d_{1}(\tilde{x})$ and $d_{2}(\tilde{x})$ have the same sign, $\tilde{x} \geq 0$, and $\beta (\tilde{x}) \neq 0$, then this implies that $c_{0}(\tilde{x}) = c_{1}(\tilde{x}) = c_{2}(\tilde{x}) = 0$, and similar to before we arrive at a contradiction. It follows that $\beta (\tilde{x}) \neq 0$ and $\kappa (\tilde{x}) > 0$. Finally, direct calculation shows that $F_{0}(c,d)(\tilde{x}) = -(\beta (\tilde{x}))^{4}(\kappa (\tilde{x}))^{3}F_0(a,b) < 0$, and from Lemma \ref{lem:bqec} we conclude that $c_{0}(\tilde{x})d_{0}(\tilde{x}), d_{2}(\tilde{x})\lambda_{1}(\tilde{x}), \gamma_{2}(\tilde{x})d_{0}(\tilde{x})d_{2}(\tilde{x}) \geq 0$; $d_{0}(\tilde{x}), d_{2}(\tilde{x}) \neq 0$; and $\gamma_{1}(\tilde{x}) > 0$. This proves that (C1) $\Rightarrow$ \ref{nl:bcca1}.

We next let condition \ref{nl:bcca1} hold, and we show that condition (C1) is satisfied. Recall from before that $F_{0}(c,d)(\tilde{x}) = -(\beta (\tilde{x}))^{4}(\kappa (\tilde{x}))^{3}F_0(a,b)$. Thus, $F_{0}(c,d)(\tilde{x}) < 0$, and from Lemma \ref{lem:bqec} we find that $c_{0}(\tilde{x}), c_{1}(\tilde{x}), c_{2}(\tilde{x}), d_{0}(\tilde{x}), d_{1}(\tilde{x}), d_{2}(\tilde{x})$ and $\lambda_{1}(\tilde{x})$ have the same sign and $\gamma_{2}(\tilde{x}) \geq 0$. Next, we recall again the relationships (\ref{eq:cbezr}) and (\ref{eq:dbezr}) and we note that $\kappa(\tilde{x}) = c(-\tilde{x})$. From (\ref{eq:dbezr}), $\beta(\tilde{x})\kappa(\tilde{x})b(s) = \beta(\tilde{x})^{2}c(s) + d(s)(s+\tilde{x})$. Since $c_{0}(\tilde{x}), c_{1}(\tilde{x}), c_{2}(\tilde{x}), d_{0}(\tilde{x}), d_{1}(\tilde{x})$ and $d_{2}(\tilde{x})$ have the same sign, $\beta(\tilde{x}) \neq 0$, $\kappa(\tilde{x}) > 0$, and $b_{i} > 0$ for at least one value of $i \in 0, \ldots , 3$, then it follows that $\beta (\tilde{x})$ has the same sign as $c_{0}(\tilde{x}), c_{1}(\tilde{x}), c_{2}(\tilde{x}), d_{0}(\tilde{x}), d_{1}(\tilde{x})$ and $d_{2}(\tilde{x})$, and $b_{i} \geq 0$ for $i = 0, 1, 2, 3$. Since, in addition, $a(s)\beta(\tilde{x}) = c(s)(s+\tilde{x}) + b(s)\alpha(\tilde{x})$, and $\alpha(\tilde{x})$ and $\beta (\tilde{x})$ have the same sign, then it follows that $a_{i} \geq 0$ for $i = 0, 1, 2, 3$. This proves that \ref{nl:bcca1} $\Rightarrow$ (C1).

A similar argument proves (C2) $\iff$ \ref{nl:bcca2}, (C3) $\iff$ \ref{nl:bcca3}, and (C4) $\iff$ \ref{nl:bcca4}.
\end{IEEEproof}

We are now in a position to prove Theorem \ref{thm:bcn1013}.
\begin{IEEEproof}[Proof of Theorem \ref{thm:bcn1013}]
We first show condition \ref{nl:bcev1}. Accordingly, let $\tilde{x} \geq 0$ be such that condition (C1) holds. It then follows from Lemma \ref{lem:bciec} that condition \ref{nl:bcca1} of that lemma holds. In particular, $\beta (\tilde{x}) \neq 0$; $c_{0}(\tilde{x})d_{0}(\tilde{x})$, $d_{2}(\tilde{x})\lambda_{1}(\tilde{x})$, $\gamma_{2}(\tilde{x})d_{0}(\tilde{x})d_{2}(\tilde{x}) \geq 0$; $d_{0}(\tilde{x}), d_{2}(\tilde{x}) \neq 0$; and $\kappa (\tilde{x}), \gamma_{1}(\tilde{x}) > 0$. It then follows that the element values $R_1, R_2, G_3, R_4, R_5, C_1, L_1$ and $L_2$ as defined in condition \ref{nl:bcev1} are all real and non-negative. Direct calculation then verifies that, with the element values thus defined, the impedance of $N_{10}$ is equal to $a/b$. Conditions \ref{nl:bcev2}--\ref{nl:bcev4} can then be shown in a similar manner.

It remains to show that if $Z$ is the impedance of a network from one of the classes $\mathcal{N}_{10}$--$\mathcal{N}_{13}$, then there exists $\tilde{x} \geq 0$ such that one of the pairs of conditions \ref{nl:bcrev1}--\ref{nl:bcrev4} holds. Accordingly, let $Z$ be the impedance of a network $N$ from the class $\mathcal{N}_{10}$ (resp., $\mathcal{N}_{11}$, $\mathcal{N}_{12}$, $\mathcal{N}_{13}$). It follows from Lemma \ref{lem:bcrc1} that there exists $\tilde{x} \geq 0$ such that the element values in $N$ take the form indicated in condition \ref{nl:bcev1} (resp., \ref{nl:bcev2}, \ref{nl:bcev3}, \ref{nl:bcev4}) of the present theorem. Since all of the element values are necessarily real and non-negative and $C_1, L_1, L_2 > 0$, then condition \ref{nl:bcca1} (resp., \ref{nl:bcca2}, \ref{nl:bcca3}, \ref{nl:bcca4}) of Lemma \ref{lem:bciec} must hold. Thus, by that same lemma, we conclude that condition (C1) (resp., (C2), (C3), (C4)) of the present theorem must also hold.

Now, let $Z$ be the impedance of a network from one of the classes $\mathcal{N}_{10}$--$\mathcal{N}_{13}$. We have shown that there exists an $\tilde{x} \geq 0$ such that one of conditions (C1)--(C4) holds. We denote this $\tilde{x}$ by $\tilde{x}_{0}$, and we will show that there exists $0 \leq \tilde{y} \leq \tilde{x}_{0}$ such that at least one of conditions (C1)--(C4) holds for all $\tilde{y} \leq \tilde{x} \leq \tilde{x}_{0}$ and at least one of the pairs of conditions \ref{nl:bcrev1}--\ref{nl:bcrev4} holds when $\tilde{x} = \tilde{y}$.

First, we note that $\alpha, \beta, \kappa$, $c_{0}$--$c_{2}$, and $d_{0}$--$d_{2}$ as defined in row (A2) of Table \ref{tab:term} are all polynomial (hence continuous) functions, as are $\gamma_{1}$--$\gamma_{3}$ and $\lambda_{1}$--$\lambda_{4}$ as defined in row (A1) of that table. It follows that either one of conditions (C1)--(C4) holds for all $\tilde{x} \geq 0$ (in which case $\tilde{x} = 0$ is such that either (C1) and (E1)(i), (C2) and (E2)(i), (C3) and (E3)(i), or (C4) and (E4)(i) holds); or there exists $0 < \tilde{y} \leq \tilde{x}_{0}$ and $\tilde{\epsilon} > 0$ such that one of conditions (C1)--(C4) holds for all $\tilde{y} \leq \tilde{x} \leq \tilde{x}_{0}$ and none of these conditions hold in the interval $\tilde{y}-\tilde{\epsilon} < \tilde{x} < \tilde{y}$. In the latter case, it follows that one of conditions \ref{nl:bcca1}--\ref{nl:bcca4} of Lemma \ref{lem:bciec} must hold for $\tilde{x} = \tilde{y}$, with one of the (non-strict) inequalities being satisfied with equality (here, we recall from Lemma \ref{lem:bciec} that conditions \ref{nl:bcca1}--\ref{nl:bcca4} of that lemma are equivalent to conditions (C1)--(C4) of the present theorem). In other words, for $\tilde{x} = \tilde{y}$, either one of conditions \ref{nl:bcrev1}--\ref{nl:bcrev4} of the present theorem holds, or $\gamma_{2}(\tilde{y}) = 0$, or $\alpha(\tilde{y}) = 0$.

Next, we note that
\begin{equation*}
\frac{\kappa}{\beta^{2}} = -\frac{d}{du}\left(\frac{\alpha}{\beta}\right).
\end{equation*}
Since $\kappa (\tilde{x})/((\beta (\tilde{x}))^{2}) > 0$ for all $\tilde{y} \leq \tilde{x} \leq \tilde{x}_{0}$, then $\alpha /\beta $ is a decreasing function in this interval. Thus, $\alpha(\tilde{y}) \neq 0$, and we conclude that either one of conditions \ref{nl:bcrev1}--\ref{nl:bcrev4} of the present theorem holds or $\gamma_{2}(\tilde{y}) = 0$.

Now, suppose that condition (C1) holds at $\tilde{x} = \tilde{y}$ but none of conditions \ref{nl:bcrev1}--\ref{nl:bcrev4} in the present theorem hold. Then $\gamma_{2}(\tilde{y}) = 0$. Since condition (C1) holds at $\tilde{x} = \tilde{y}$ but $\lambda_{1}(\tilde{y}), c_{0}(\tilde{y}) \neq 0$, then from Lemma \ref{lem:bciec} it follows that condition \ref{nl:bcca1} of that lemma holds and $c_{0}(\tilde{y})d_{0}(\tilde{y}), d_{2}(\tilde{y})\lambda_{1}(\tilde{y}), \gamma_{1}(\tilde{y}) > 0$. Also, since $\gamma_{2}(\tilde{y}) = 0$, then $F_{0}(c,d)(\tilde{y}) = \gamma_{1}(\tilde{y})\gamma_{3}(\tilde{y}) < 0$, and it follows that $\gamma_{3}(\tilde{y}) < 0$. Moreover, $\lambda_{1}(\tilde{y}) = d_{1}(\tilde{y})\gamma_{1}(\tilde{y})$ and $\lambda_{3}(\tilde{y}) = -d_{1}(\tilde{y})\gamma_{3}(\tilde{y})$, and hence $\lambda_{3}(\tilde{y})$ has the same sign as $d_{1}(\tilde{y})$. We conclude that condition (C2) also holds at $\tilde{x} = \tilde{y}$. Since, in addition, condition (E2) of the present theorem does not hold at $\tilde{x} = \tilde{y}$, then from Lemma \ref{lem:bciec} it follows that condition \ref{nl:bcca2} of that lemma holds and $c_{2}(\tilde{y})d_{2}(\tilde{y}), d_{0}(\tilde{y})\lambda_{3}(\tilde{y}), -\gamma_{3}(\tilde{y}) > 0$. Finally, since $\alpha (\tilde{y}) \neq 0$, and either $\gamma_{2}(\tilde{x})d_{0}(\tilde{x})d_{2}(\tilde{x}) \geq 0$ or $\gamma_{2}(\tilde{x})d_{0}(\tilde{x})d_{2}(\tilde{x}) \leq 0$ for all $\tilde{x} \in \mathbb{R}$, then it follows that there exists $\tilde{\rho} > 0$ such that either condition \ref{nl:bcca1} or condition \ref{nl:bcca2} of Lemma \ref{lem:bciec} holds for all $\tilde{y} - \tilde{\rho} < \tilde{x} < \tilde{y}$. It follows from Lemma \ref{lem:bciec} that either condition (C1) or condition (C2) hold for all $\tilde{y} - \tilde{\rho} < \tilde{x} < \tilde{y}$, which contradicts the assumption that none of conditions (C1)--(C4) hold in the interval $\tilde{y} - \tilde{\epsilon} < \tilde{x} < \tilde{y}$. We conclude that, if condition (C1) holds at $\tilde{x} = \tilde{y}$, then one of conditions \ref{nl:bcrev1}--\ref{nl:bcrev4} in the present theorem hold. A similar argument then shows that, if condition (C2) (resp., (C3), (C4)) holds at $\tilde{x} = \tilde{y}$, then one of conditions \ref{nl:bcrev1}--\ref{nl:bcrev4} in the present theorem hold, which completes the proof.
\end{IEEEproof}

Theorems \ref{thm:bcn14}--\ref{thm:bqn14} can be proved in a similar manner to the above proof of Theorem \ref{thm:bcn1013}. In the case of Theorems \ref{thm:bqn1013} and \ref{thm:bqn14}, the parameter $\tilde{z}$ is varied as opposed to $\tilde{x}$, so $c_{0}$--$c_{2}$ and $\tilde{x}$ are unchanged, whereupon we arrive at the equality constraints (F1)--(F5).

We next prove Theorem \ref{thm:mr3re}.
\begin{IEEEproof}[Proof of Theorem \ref{thm:mr3re}]
That $Z$ is the impedance of a network from one of the network classes listed in the present theorem follows from Theorems \ref{thm:z12nmgs}--\ref{thm:z21nmgs} and Lemmas \ref{lem:z30mgs}--\ref{lem:z03mgs}. Also, if $Z$ is the impedance of a network from one of the classes $\mathcal{N}_1$--$\mathcal{N}_3$ (resp., $\mathcal{N}_4$--$\mathcal{N}_9$, $\mathcal{N}_{15}$) then an explicit series-parallel generic network realization for $Z$ can be obtained from Lemma \ref{lem:blgs} (resp., \ref{lem:bqrrc}, \ref{lem:3resc}).

Now, let $Z$ be the impedance of a network from one of the classes $\mathcal{N}_{10}$--$\mathcal{N}_{13}$. If $Z$ is bilinear or constant, then $Z \in \mathcal{Z}_1$ by Lemma \ref{lem:blec}, in which case an explicit realization for $Z$ is obtained from Lemma \ref{lem:blgs}. Accordingly, it remains to consider the cases in which $Z$ is biquadratic and bicubic. If $Z$ is bicubic, then $Z$ is the impedance of a network from one of the classes $\mathcal{N}_{16}$--$\mathcal{N}_{27}$ by Theorem \ref{thm:bcn1013}, and an explicit series-parallel generic network realization for $Z$ is also obtained from that theorem. If, on the other hand, $Z$ is biquadratic, then it follows from Theorem \ref{thm:bqn1013} that either (i) $Z \in \mathcal{Z}_{0,2} \cup \mathcal{Z}_{1,1}$, or (ii) $Z$ is the impedance of a network from one of the classes $\mathcal{N}_{17}$, $\mathcal{N}_{20}$, $\mathcal{N}_{22}$, $\mathcal{N}_{23}$, $\mathcal{N}_{25}$ or $\mathcal{N}_{26}$. In case (i) (resp., (ii)), an explicit series-parallel generic network realization for $Z$ is obtained from Lemma \ref{lem:bqrrc} (resp., Theorem \ref{thm:bqn1013}). 

Explicit realizations for the case in which $Z$ is the impedance of a network from $\mathcal{N}_{14}$ can be obtained similarly using Theorems \ref{thm:bcn14} and  \ref{thm:bqn14}. Finally, explicit realizations for the cases in which $Z$ is the impedance of a network from $\mathcal{N}_{10}^i$--$\mathcal{N}_{14}^i$, $\mathcal{N}_{10}^p$--$\mathcal{N}_{14}^p$, $\mathcal{N}_{10}^d$--$\mathcal{N}_{14}^d$ or $\mathcal{N}_{15}^i$ are routinely obtained using the results of Section \ref{sec:ncm}.
\end{IEEEproof}

We finish this section by proving Theorem \ref{thm:z12mgs}, and Theorem \ref{thm:z21mgs} can be proved similarly.
\begin{IEEEproof}[Proof of Theorem \ref{thm:z12mgs}]
If $Z$ is the impedance of a network from one of the classes $\mathcal{N}_{10}$--$\mathcal{N}_{14}$, then it follows from Lemmas \ref{lem:blgs}--\ref{lem:bqrrc} and Theorems \ref{thm:bcn1013}--\ref{thm:bqn14} that $Z$ is also the impedance of a network from one of the classes $\mathcal{N}_{1}$--$\mathcal{N}_{3}$, $\mathcal{N}_{5}$--$\mathcal{N}_{9}$, or $\mathcal{N}_{16}$--$\mathcal{N}_{30}$. Then, using the results of Section \ref{sec:ncm}, it follows that if $Z$ is the impedance of a network from one of the classes $\mathcal{N}_{10}^p$--$\mathcal{N}_{14}^p$, then $Z$ is also the impedance of a network from one of the classes $\mathcal{N}_{1}$--$\mathcal{N}_{3}$, $\mathcal{N}_{5}$--$\mathcal{N}_{9}$, or $\mathcal{N}_{16}^p$--$\mathcal{N}_{30}^p$. It follows from Theorem \ref{thm:z12nmgs} that the union of $\mathcal{N}_{1}$--$\mathcal{N}_{3}$, $\mathcal{N}_{5}$--$\mathcal{N}_{9}$, $\mathcal{N}_{16}$--$\mathcal{N}_{30}$ and $\mathcal{N}_{16}^p$--$\mathcal{N}_{30}^p$ is a generating set for $\mathcal{Z}_{1,2}$. It is also straightforward to verify that all of the networks in this generating set are generic, so the generating set is minimal. 

To complete the proof of the present theorem, we will show that if $Z$ is the impedance of a network from one of the classes $\mathcal{N}_{16}$, $\mathcal{N}_{19}$, $\mathcal{N}_{28}$, $\mathcal{N}_{16}^p$, $\mathcal{N}_{19}^p$ or $\mathcal{N}_{28}^p$, then $Z$ is also the impedance of a network from one of the classes listed in the present theorem statement. To see this, note initially that networks from the classes $\mathcal{N}_{16}$ and $\mathcal{N}_{28}$ each contain a subnetwork comprising one inductor, one capacitor and three resistors. From Lemmas \ref{lem:blgs}--\ref{lem:bqrrc}, the impedance of this subnetwork is also realized by a network from one of the classes $\mathcal{N}_{1}$--$\mathcal{N}_{3}$ or $\mathcal{N}_{6}$--$\mathcal{N}_{9}$, and it can then be verified that the impedance of the overall network is also realized by a network from one of the classes $\mathcal{N}_{1}$--$\mathcal{N}_{3}$, $\mathcal{N}_{5}$--$\mathcal{N}_{9}$, $\mathcal{N}_{18}$, $\mathcal{N}_{21}$, $\mathcal{N}_{24}$ or $\mathcal{N}_{27}$. Similarly, networks from the class $\mathcal{N}_{19}$ contain a subnetwork comprising two inductors and three resistors. From Lemmas \ref{lem:blgs}--\ref{lem:bqrrc}, if follows that the impedance of this subnetwork is also realized by a network from one of the classes $\mathcal{N}_{1}$, $\mathcal{N}_{3}$ or $\mathcal{N}_{5}$, and it follows that the impedance of the overall network is also realized by a network from one of the classes $\mathcal{N}_{1}$--$\mathcal{N}_{3}$, $\mathcal{N}_{5}$--$\mathcal{N}_{9}$ or $\mathcal{N}_{30}$. A similar argument then covers the cases in which $Z$ is the impedance of a network from one of the classes $\mathcal{N}_{16}^p$, $\mathcal{N}_{19}^p$ or $\mathcal{N}_{28}^p$, and completes the proof.
\end{IEEEproof}

\section{Passive mechanical control}
\label{sec:pmc}

\begin{figure}[!t]
\scriptsize
  \begin{center}
    \leavevmode
    \includegraphics[page=2, width=1.0\hsize]{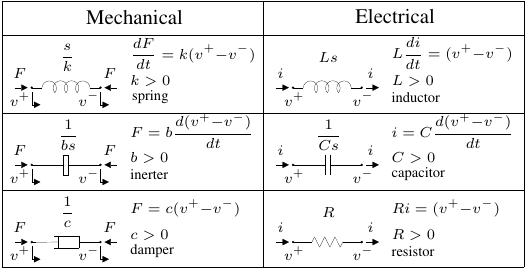}
\end{center}
\caption{Passive electrical and mechanical elements.}
\label{fig:ema}
\end{figure}

The relevance of this paper to passive mechanical control is due to an analogy between electrical and mechanical networks. This so-called force-current analogy is outlined in Fig.\ \ref{fig:ema}. Indeed, it was the absence of a mechanical equivalent to a capacitor that inspired the invention of the inerter \cite{mcs02}, a two-terminal passive device for which the force transmitted through the device is proportional to the relative acceleration of the two terminals (in contrast, a mass has the property that the force applied is proportional to its acceleration \emph{relative to a fixed point}, which is analogous to a \emph{grounded} capacitor). The force-current analogy establishes a one-to-one correspondence between electrical and mechanical networks, whereby resistors are replaced with dampers, inductors with springs, and capacitors with inerters. With these substitutions, the impedances of the corresponding networks are equal.\footnote{Here, we define the impedance $Z(s)$ as the transfer function from driving-point current (resp., transmitted force) to driving-point voltage (resp., relative velocity of the two terminals), and the admittance $Y(s) = 1/Z(s)$. Note that the opposite convention is sometimes followed in the mechanical case.\label{fn:ad}} Thus, the results in this paper provide statements about minimal mechanical network realizations.

To illustrate the practical relevance of our results, we now consider a specific example. The paper \cite{fucheng_9b} considered the optimal design for a passsive train suspension system. The suspension system is shown in Fig.\ \ref{fig:mnc}, where $m_s, m_b$ and $m_w$ represent the masses of the train body, bogie and wheel; and $k_w$ and $c_w$ represent the stiffness and damping coefficient for the train wheel. The objective of the design is to obtain suspension admittances\textsuperscript{\ref{fn:ad}} $Q_1$ and $Q_2$ in order to optimise two measures of the system's performance: passenger comfort and dynamic wheel load. The case of optimising passenger comfort amounts to minimising the $H_2$ norm of the transfer function from $z_r$ to $\tfrac{dz_s}{dt}$. In one investigation, the admittance $Q_1(s) = k_s/s + K_1(s)$ was chosen to optimise this metric for the case in which the train parameters were as indicated in Fig.\ \ref{fig:mnc}, where $K_1(s)$ is a positive-real function whose McMillan degree does not exceed three. This resulted in the optimal admittance\textsuperscript{\ref{fn:ad}}
\begin{equation}
K_1(s) = \frac{3440s^3 + 84150s^2 + 190750s + 2010}{s^3 + 9.2741s^2 + 59.6319s + 134.6736}.\label{eq:k1d}
\end{equation}
From Theorem \ref{thm:mr3re} of the present paper, we obtain six realizations for the admittance $K_1(s)$: two realizations from the class $\mathcal{N}_{19}^p$; two from the class $\mathcal{N}_{20}^p$; one from the class $\mathcal{N}_{29}^p$; and one from the class $\mathcal{N}_{30}^p$. One of the mechanical networks that realizes the admittance $K_1(s)$ is shown in Fig.\ \ref{fig:mncrel} (this is the analogue of one of the electrical realizations from $\mathcal{N}_{20}^p$).

In a second investigation in \cite{fucheng_9b}, it is shown that the passenger comfort metric can be improved by a further $3.6\%$ if, in addition, $Q_2(s) = k_b/s + K_2(s)$ where $K_2(s)$ is also chosen from among the entire class of positive-real functions whose McMillan degree does not exceed three. However, Theorems \ref{thm:mr3re}--\ref{thm:bqn14} of the present paper can also be used to show that the functions $K_1(s)$ and $K_2(s)$ thus obtained cannot be realized by a series-parallel network containing at most three energy storage elements.

Similar optimization problems can be posed for a wide range of other applications (see, e.g., \cite{chen_14, fucheng_10, fucheng_12, jiang_vsd_12, fucheng_07, LNW_SVS, Limebeer_steering2, Limebeer_Steering, YJN_MLGSS}), but are often very computationally demanding. As we argue in the following two paragraphs, the computational complexity of these design problems can be substantially reduced by fundamental studies of the type presented in this paper.

The paper \cite{ZJ_SIA} noted two approaches to an optimization problem of this type---the \emph{immittance-based} and \emph{structure-based} approaches---and also proposed a third hybrid method termed the \emph{structure-immittance approach}. In the immittance-based approach, the optimization is performed over all controller impedances taken from a certain class. Here, a key challenge is to identify an appropriate class of impedances. If this is defined too broadly, then there will be no guarantee that the impedance can be realized by a passive mechanical system. In many studies (e.g., \cite{smith_06}), the optimization is performed over the class of positive-real impedances (with an upper bound on the McMillan degree), in which case it is known that a passive mechanical realization necessarily exists. But the number of elements required to realize the optimal impedance cannot be predicted in advance, and is potentially large. For example, to realize a biquadratic \emph{minimum} function with a series-parallel network, six energy storage elements are required \cite{HugSmSP}; yet to realize a biquadratic \emph{regular} function requires only two energy storage elements \cite{JiangSmith11}. Thus, instead of optimizing over the class of positive-real impedances, it is preferable to optimize over the class of impedances that can be realized by networks of a given complexity. One contribution of this paper is to describe one such class of impedances, namely the class of impedances that are realized by series-parallel networks containing three energy storage elements (springs and inerters) and a finite number of resistive elements (dampers). Also, in cases where an optimization procedure has already been carried out, the results of this paper allow one to determine whether the resulting immittance can be realized by a series-parallel network containing three or fewer energy storage elements. If a realization exists, our results also provide an explicit series-parallel generic network realization.

\tikzset{ mn/.pic={
\ctikzset{bipoles/thickness=1}
\node[anchor=center] at (1,0) (O) {};
\draw (O) to[short] ++(4.0,0)
($(O)+(0.5,0)$) to[spring] ++(0,3)
($(O)+(3.5,0)$) to[twoport] 
++(0,2.0)
to[spring] ++(0,1.0)
($(O)+(0,3.0)$) to[short] ++(4.0,0)
to[short] ++(0,0.3)
to[short] ++(-4.0,0)
to[short] ++(0,-0.3)
($(O)+(0.5,3.3)$) to[damper] ++(0,1.0)
($(O)+(0.5,5.3)$) to[spring] ++(0,1.0)
($(O)+(0,6.3)$) to[short] ++(4.0,0)
to[short] ++(0,0.3)
to[short] ++(-4.0,0)
to[short] ++(0,-0.3)
($(O)+(0.5,6.6)$) to[spring] ++(0,1.0)
($(O)+(0.5,8.6)$) to[spring] ++(0,1.0)
($(O)+(0,9.6)$) to[short] ++(4.0,0)
to[short] ++(0,0.3)
to[short] ++(-4.0,0)
to[short] ++(0,-0.3);}}

\tikzset{ mn2/.pic={
\ctikzset{bipoles/thickness=1}
\node[anchor=center] at (1,0) (O) {};
\draw (O) to[short] ++(3.0,0)
($(O)+(0.5,0)$) to[spring=$k_w$] ++(0,2.0)
($(O)+(2.5,2.0)$) to[damper=$c_w$] 
++(0,-2.0)
($(O)+(0,2.0)$) to[short] ++(3.0,0)
to[short] ++(0,1.5)
to[short] ++(-3.0,0)
to[short] ++(0,-1.5)
($(O)+(1.5,3.5)$) to[twoport, t={\Large $Q_2$}] ++(0,2.0)
($(O)+(0,5.5)$) to[short] ++(3.0,0)
to[short] ++(0,1.5)
to[short] ++(-3.0,0)
to[short] ++(0,-1.5)
($(O)+(1.5,7.0)$) to[twoport, t={\Large $Q_1$}] ++(0,2.0)
($(O)+(0,9.0)$) to[short] ++(3.0,0)
to[short] ++(0,1.5)
to[short] ++(-3.0,0)
to[short] ++(0,-1.5)
($(O)+(4.0,0)$) to[short] ++(0.8,0)
to[short] ++(0,0.8)
to[short,i_>={\Large $z_r$}] ++(0,0.2)
($(O)+(4.0,9.75)$) to[short] ++(0.8,0)
to[short] ++(0,0.8)
to[short,i_>={\Large $z_s$}] ++(0,0.2);}}

\begin{figure}[!b]
\centering
\normalsize
\begin{tikzpicture}[scale=0.57, every node/.style={transform shape}, font=\Large]
\pic at (0,0) [] {mn2};
\node at (2.5,2.8) {\Large $m_w$};
\node at (2.5,6.3) {\Large $m_b$};
\node at (2.5,9.8) {\Large $m_s$};
\node at (10.0,6.0) {\Large $\begin{array}{l}m_s = 3500\mathrm{kg} \\ m_b = 250\mathrm{kg} \\ m_w = 350\mathrm{kg} \\ k_w = 8{\times}10^9 \mathrm{N/m} \\ c_w = 670{\times}10^3 \mathrm{Ns/m} \\ Q_1(s) = k_s/s + K_1(s) \\ Q_2(s) = k_b/s + c_b \\ k_s = 141{\times} 10^3 \mathrm{N/m} \\ k_b = 1260{\times}10^3 \mathrm{N/m} \\ c_b = 7100 \mathrm{Ns/m}\end{array}$};
\end{tikzpicture}
\caption{One wheel train model (see \cite{fucheng_9b}).}
\label{fig:mnc}
\end{figure}
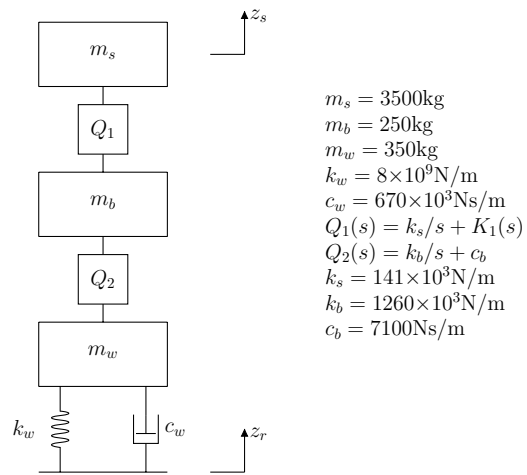

\tikzset{ mnrel/.pic={
\ctikzset{bipoles/thickness=1}
\node[anchor=center] at (1,0) (O) {};
\draw (O) to[short,*-] ++(0.3,0)
to[short] ++(0,1.4)
to[damper=$\frac{1}{c_3}$] ++(5.4,0)
to[short] ++(0,-1.4)
to[short,-*] ++(0.3,0)
($(O)+(0.3,0)$) to[damper=$\frac{1}{c_4}$] ++(1.7,0)
to[short] ++(0,-1.4)
to[spring=$\frac{s}{k_2}$] ++(-1.7,0)
to[short] ++(0,1.4)
($(O)+(2.0,-0.7)$) to[short] ++(0.3,0)
to[short] ++(0,0.7)
to[damper=$\frac{1}{c_1}$] ++(3.4,0)
to[short] ++(0,-2.8)
to[spring=$\tfrac{s}{k_1}$] ++(-1.7,0)
to[short] ++(0,0.7)
to[short] ++(-0.7,0)
to[short] ++(0,0.5)
to[short] ++(-0.3,0)
to[short] ++(0,-0.5)
to[short] ++(-0.7,0)
to[short] ++(0,1.4)
($(O)+(3.0,-2.1)$) to[short] ++(0,-0.5)
to[short] ++(0.3,0)
to[short] ++(0,0.5)
($(O)+(4.0,-2.1)$) to[short] ++(0,0.7)
to[damper=$\frac{1}{c_2}$] ++(1.7,0);}}

\begin{figure}[!b]
\centering
\normalsize
\begin{tikzpicture}[scale=0.65, every node/.style={transform shape}, font=\large]
\pic at (0,0) [] {mnrel};
\node at (4.1,-1.3) {\large $\frac{1}{b_1s}$};
\node at (10.0,0) {\large $\begin{array}{l} c_1 = 5.7587\mathrm{Ns/m} \\
c_2 = 4405.3\mathrm{Ns/m} \\
c_3 = 9.1662\mathrm{Ns/m} \\
c_4 = 15439\mathrm{Ns/m} \\
k_1 = 96960\mathrm{N/m} \\
k_2 = 38869\mathrm{N/m} \\
b_1 = 1409.8\mathrm{kg} \end{array}$};
\end{tikzpicture}
\caption{Mechanical realization of the admittance $K_1(s)$ in (\ref{eq:k1d}).}
\label{fig:mncrel}
\end{figure}
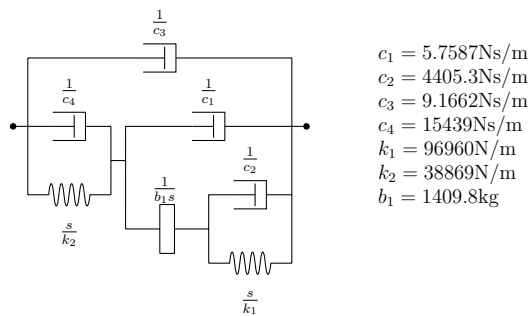

In the structure-based approach, the controller's configuration is fixed, and the element parameters are varied in order to optimize performance. This approach has the advantage of providing a guarantee on the controller's complexity. However, the approach fails to identify alternative (potentially simpler) controller configurations with improved performance. To overcome this drawback, it is possible to perform structure-based optimization on a number of different configurations. Indeed, the structure-immittance approach of \cite{ZJ_SIA} outlines a systematic way of considering all possible configurations of series-parallel networks that have a given number of each type of element. However, except in particularly simple cases, it is unlikely to be computationally feasible to cover all configurations, since the number of series-parallel networks grows significantly as the number of elements is increased \cite{Lomnicki_SP}. Another contribution of this paper is to show that the impedance of any series-parallel network containing three energy storage elements and a finite number of resistors (dampers) can always be realized by one of a small number of series-parallel network configurations, each of which contains at most seven elements. Nevertheless, finding an optimal network from amongst this class of networks for a given application remains a challenging non-convex and high dimensional optimization problem and is a topic for further research.

\section{Conclusions}
In this paper, we developed a novel continuity-based argument to solve the minimal network realization problem for the class of impedance functions realized by series-parallel networks containing three energy storage elements and a finite number of resistors. Specifically, it is shown that any such impedance is realized by a network from one of the classes described in Lemmas \ref{lem:z30mgs}--\ref{lem:z03mgs} and Theorems \ref{thm:z12mgs}--\ref{thm:z21mgs}, and explicit formulae are provided for the element values in terms of the coefficents in the impedance function. As outlined in Section \ref{sec:pmc}, this represents an important contribution towards the design of passive mechanical controllers containing the inerter. Finally, we note that the mathematical structures encountered in electric circuits are also prevalent in other physical systems, such as multibody systems, hydraulic networks, chemical reaction networks and power systems; and similar network dynamics arise in other fields, such as consensus and clustering algorithms \cite{VDS_PHS}. The adaptation and extension of these results to the design of such systems is a topic for future research.

\bibliographystyle{IEEEtran}
\bibliography{IEEEabrv,bicubic_rlc_refs}
\begin{IEEEbiography}[{\includegraphics[width=1in, height=1.25in, clip, keepaspectratio]{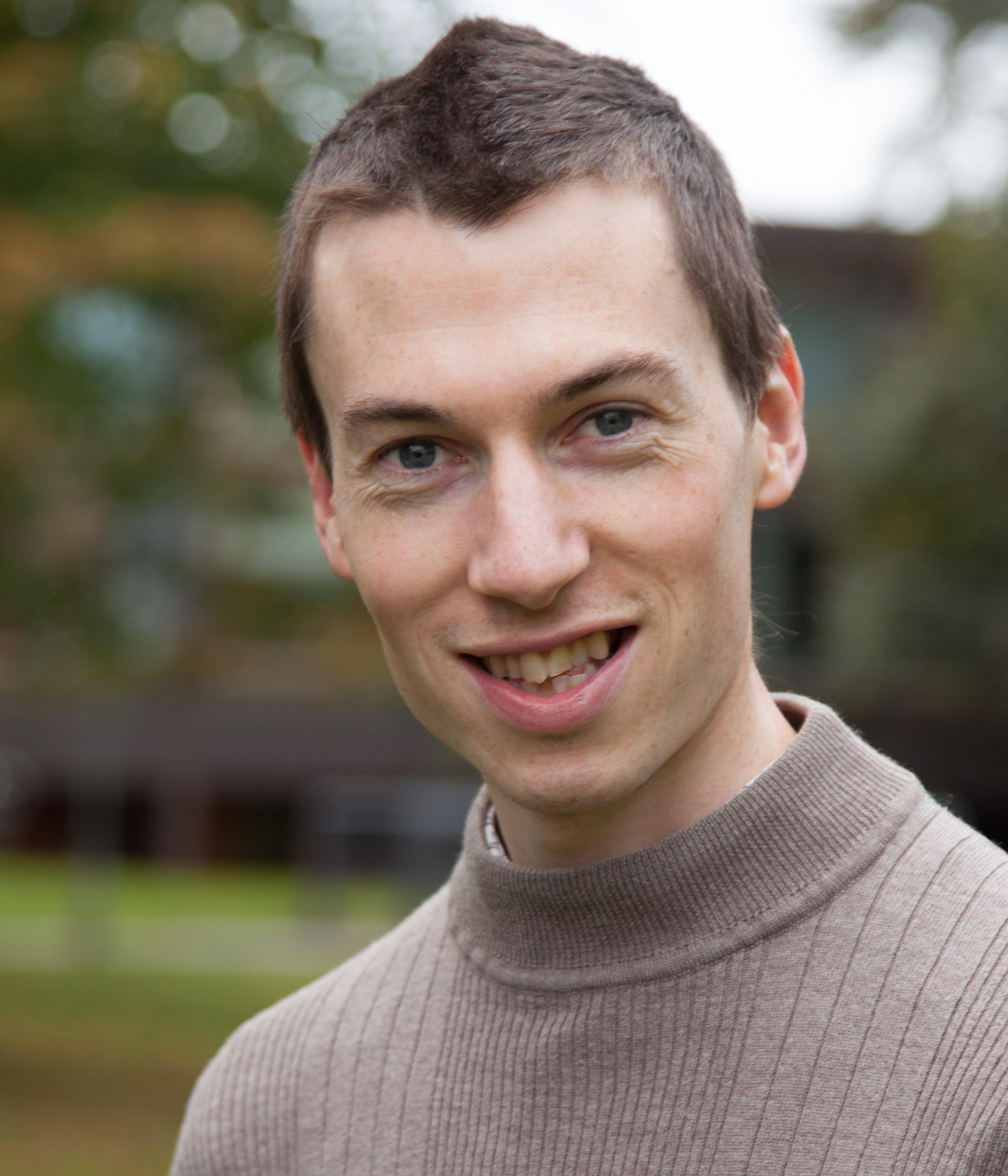}}]{Timothy H. Hughes}
received the M.Eng. degree in mechanical engineering, and the Ph.D. degree in control engineering, from the University of Cambridge, U.K., in 2007 and 2014, respectively.

From 2007 to 2010 he was employed as a Mechanical Engineer at The Technology Partnership, Hertfordshire, U.K; and from 2013 to 2017 he held a Henslow Research Fellowship at the University of Cambridge. He is now a Lecturer in the Department of Mathematics at the University of Exeter.

 \end{IEEEbiography}
\vfill
\pagebreak
\appendix
\begin{table}[!h]
\begin{tabular}{|p{0.6cm}|p{7.6cm}|}
\hline (A1)& {$ \displaystyle \begin{aligned}
\gamma_{1} &= c_{1}d_{0} - c_{0}d_{1}, \\
\gamma_{2} &= c_{2}d_{0} - c_{0}d_{2}, \\
\gamma_{3} &= c_{2}d_{1} - c_{1}d_{2}, \\
\lambda_{1} &= d_{1}\gamma_{1} - d_{0}\gamma_{2}, \\
\lambda_{2} &= c_{1}\gamma_{3} - c_{2}\gamma_{2}, \\
\lambda_{3} &= d_{2}\gamma_{2} - d_{1}\gamma_{3}, \\
\lambda_{4} &= c_{0}\gamma_{2} - c_{1}\gamma_{1}. \end{aligned}$} \\
\hline (A2)& 
{$\displaystyle \begin{aligned}
\begin{bmatrix}c_{0}& c_{1}& c_{2}\end{bmatrix}(u) &= \begin{bmatrix}1& -u& u^{2}\end{bmatrix}\mathcal{B}(a,b),\rule{0pt}{2.6ex}  \\
\alpha (u) &= -a_3 u^3 + a_2 u^2 - a_1 u + a_0, \\
\beta(u) &= -b_3 u^3 + b_2 u^2 - b_1 u + b_0, \\
\kappa (u) &= c_2(u) u^2 -c_1(u) u + c_0(u), \\
\begin{bmatrix}d_{0}& d_{1}& d_{2}\end{bmatrix}(u) &= \beta (u)\begin{bmatrix}1& -u& u^{2}\end{bmatrix}\mathcal{B}(b,c)(u), \\ 
\text{where } c(s) &= c_{2}s^{2} + c_{1}s+ c_{0},\footnote{In the definition of $d_0$--$d_2$, $b$ and $c$ are to be viewed as polynomials in the indeterminate $s$, where the coefficients of $c$ are themselves polynomials in the indeterminate $u$. See footnote \ref{fn:bmp}.\label{fn:bmp2}}
\end{aligned}$}\\
\hline (A3)& {$\displaystyle \begin{aligned} \begin{bmatrix}c_{0}& c_{1}& c_{2}\end{bmatrix}(u) &= \begin{bmatrix}-u^{2}& u& -1\end{bmatrix}\mathcal{B}(a,b),\rule{0pt}{2.6ex} \\
\alpha (u) &= -a_{0}u^{3}+a_{1}u^{2}-a_{2}u+a_{3}, \\
\beta (u) &= -b_{0}u^{3}+b_{1}u^{2}-b_{2}u+b_{3}, \\
\kappa (u) &= c_{0}(u)u^{2} -c_{1}(u)u + c_{2}(u), \\
\begin{bmatrix}d_{0}& d_{1}& d_{2}\end{bmatrix}(u) &= \beta (u)\begin{bmatrix}-u^{2}& u& -1\end{bmatrix}\mathcal{B}(b,c)(u), \\
\text{where } c(s) &= c_{2}s^{3} + c_{1}s^{2}+ c_{0}s.\textsuperscript{\ref{fn:bmp2}}\end{aligned}$} \\ 
\hline (A4) & {$\displaystyle \begin{aligned} \alpha (u) &= a_2 u^2 - a_1 u + a_0,\rule{0pt}{2.6ex} \\ \beta (u) &= b_2 u^2 - b_1 u + b_0, \\ 
\begin{bmatrix}f_{0}& f_{1}\end{bmatrix}(u) &= \begin{bmatrix}1& -u\end{bmatrix}\mathcal{B}(a,b), \\ 
c_{2}(u) &= f_{1}(u), \\
c_{1}(u) &= f_{0}(u)+uf_{1}(u), \\ 
c_{0}(u) &= uf_{0}(u), \\
d_{2}(u,v) &= \beta (u)b_{2}, \\
d_{1}(u,v) &= \beta (u)(b_{1} - vf_{1}(u)), \\
d_{0}(u,v) &= \beta (u)(b_{0} - vf_{0}(u)), \\ 
\eta (u,v) &= b_2 u^{2} - (b_{1} - vf_{1}(u))u +  b_{0} - vf_{0}(u).\end{aligned}$} \\ 
\hline (A5)& {$\displaystyle \begin{aligned} \alpha (u) &= a_{0}u^{2} -a_{1}u + a_{2},\rule{0pt}{2.6ex} \\
\beta (u) &= b_{0}u^{2} - b_{1}u + b_{2}, \\
\begin{bmatrix}f_{0}& f_{1}\end{bmatrix}(u) &= \begin{bmatrix}-u& 1\end{bmatrix}\mathcal{B}(a,b), \\
c_{2}(u) &= uf_{1}(u), \\
c_{1}(u) &= uf_{0}(u)+f_{1}(u), \\
c_{0}(u) &= f_{0}(u), \\
d_{2}(u,v) &= \beta (u)(b_{2} - vf_{1}(u)), \\
d_{1}(u,v) &= \beta (u)(b_{1} - vf_{0}(u)), \\
d_{0}(u,v) &= \beta (u)b_{0}, \\
\eta (u,v) &= b_{0}u^{2} -(b_{1} - vf_{0}(u))u + b_{2} - vf_{1}(u).\end{aligned}$} \\ \hline
\end{tabular}
\caption{Definition of terminology}
\label{tab:term}
\end{table}

\begin{table}
\begin{tabular}{|p{0.7cm}|p{7.2cm}|}
\hline (C1)& $a_{i}, b_{i} {\geq} 0$ for $i {=} 0,1,2,3$; $\alpha (\tilde{x}), \beta (\tilde{x}), c_{0}(\tilde{x}), c_{1}(\tilde{x})$, $c_{2}(\tilde{x}), d_{0}(\tilde{x}), d_{1}(\tilde{x}), d_{2}(\tilde{x})$ and $\lambda_{1}(\tilde{x})$ have the same sign\footnote[2]{See footnote \ref{fn:ss}}; $\kappa (\tilde{x}), \gamma_{2}(\tilde{x}) \geq 0$; and $F_0(a,b) > 0$. \\ \hline
(C2)& $a_{i}, b_{i} {\geq} 0$ for $i {=} 0,1,2,3$; $\alpha (\tilde{x}), \beta (\tilde{x}), c_{0}(\tilde{x}), c_{1}(\tilde{x}),$ $c_{2}(\tilde{x}), d_{0}(\tilde{x}), d_{1}(\tilde{x}), d_{2}(\tilde{x})$ and $\lambda_{3}(\tilde{x})$ have the same sign; $\kappa (\tilde{x}), -\gamma_{2}(\tilde{x}) \geq 0$; and $F_0(a,b) > 0$. \\ \hline
(C3)& $a_{i}, b_{i} {\geq} 0$ for $i {=} 0,1,2,3$; $\alpha (\tilde{x}), \beta (\tilde{x}), c_{0}(\tilde{x}), c_{1}(\tilde{x}),$ $c_{2}(\tilde{x}), d_{0}(\tilde{x}), d_{1}(\tilde{x}), d_{2}(\tilde{x})$ and $\lambda_{4}(\tilde{x})$ have the same sign; $\kappa (\tilde{x}), -\gamma_{2}(\tilde{x}) \geq 0$; and $F_0(a,b) > 0$. \\ \hline
(C4)& $a_{i}, b_{i} {\geq} 0$ for $i {=} 0,1,2,3$; $\alpha (\tilde{x}), \beta (\tilde{x}), c_{0}(\tilde{x}), c_{1}(\tilde{x}),$ $c_{2}(\tilde{x}), d_{0}(\tilde{x}), d_{1}(\tilde{x}), d_{2}(\tilde{x})$ and $\lambda_{2}(\tilde{x})$ have the same sign; $\kappa (\tilde{x}), \gamma_{2}(\tilde{x}) \geq 0$; and $F_0(a,b) > 0$. \\ \hline
(C5)& $a_{i}, b_{i} {\geq} 0$ for $i {=} 0,1,2,3$; $\alpha (\tilde{x}), \beta (\tilde{x}), c_{0}(\tilde{x}), c_{1}(\tilde{x}),$ $c_{2}(\tilde{x}), d_{0}(\tilde{x}), d_{1}(\tilde{x}), d_{2}(\tilde{x})$ and $\lambda_{1}(\tilde{x})$ have the same sign; $\kappa (\tilde{x}), \gamma_{2}(\tilde{x}) \geq 0$; and $F_0(a,b) > 0$. \\ \hline
(Q1)& $c_{0}, c_{1}, c_{2}, d_{0}, d_{1}, d_{2}, \lambda_{1}$ have the same sign; and $\gamma_{2} \geq 0$. \\ \hline
(Q2)& $c_{0}, c_{1}, c_{2}, d_{0}, d_{1}, d_{2}, \lambda_{2}$ have the same sign; and $\gamma_{2} \geq 0$. \\ \hline
(Q3)& $c_{0}, c_{1}, c_{2}, d_{0}, d_{1}, d_{2}, \lambda_{3}$ have the same sign; and $\gamma_{2} \leq 0$. \\ \hline
(Q4)& $c_{0}, c_{1}, c_{2}, d_{0}, d_{1}, d_{2}, \lambda_{4}$ have the same sign; and $\gamma_{2} \leq 0$. \\ \hline
(Q5)& $a_{i}, b_{i} \geq 0$ for $i = 0,1,2$; $\alpha (\tilde{x}), \beta (\tilde{x}), \tilde{z}, d_{0}(\tilde{x},\tilde{z}),$ $d_{1}(\tilde{x},\tilde{z}), d_{2}(\tilde{x},\tilde{z}), f_{0}(\tilde{x}), f_{1}(\tilde{x})$ and $\lambda_{1}(\tilde{x},\tilde{z})$ have the same sign; $\gamma_{2}(\tilde{x},\tilde{z}), -\beta (\tilde{x})\eta (\tilde{x},\tilde{z})F_0(a,b) \geq 0$; and $\eta (\tilde{x},\tilde{z}), F_0(a,b) \neq 0$. \\ \hline
(Q6)& $a_{i}, b_{i} \geq 0$ for $i = 0,1,2$; $\alpha (\tilde{x}), \beta (\tilde{x}), \tilde{z}, d_{0}(\tilde{x},\tilde{z}),$ $d_{1}(\tilde{x},\tilde{z}), d_{2}(\tilde{x},\tilde{z}), f_{0}(\tilde{x}), f_{1}(\tilde{x})$ and $\lambda_{3}(\tilde{x},\tilde{z})$ have the same sign; $\gamma_{2}(\tilde{x},\tilde{z}), \beta (\tilde{x})\eta (\tilde{x},\tilde{z})F_0(a,b) \leq 0$; and $\eta (\tilde{x},\tilde{z}), F_0(a,b) \neq 0$ \\ \hline
(Q7)& $a_{i}, b_{i} \geq 0$ for $i = 0,1,2$; $\alpha (\tilde{x}), \beta (\tilde{x}), \tilde{z}, d_{0}(\tilde{x},\tilde{z}),$ $d_{1}(\tilde{x},\tilde{z}), d_{2}(\tilde{x},\tilde{z}), f_{0}(\tilde{x}), f_{1}(\tilde{x})$ and $\lambda_{4}(\tilde{x},\tilde{z})$ have the same sign; $\gamma_{2}(\tilde{x},\tilde{z}), \beta (\tilde{x})\eta (\tilde{x},\tilde{z})F_0(a,b) \leq 0$; and $\eta (\tilde{x},\tilde{z}), F_0(a,b) \neq 0$. \\ \hline
(Q8)& $a_{i}, b_{i} \geq 0$ for $i {=} 0,1,2$; $\alpha (\tilde{x}), \beta (\tilde{x}), \tilde{z}, d_{0}(\tilde{x},\tilde{z}),$ $d_{1}(\tilde{x},\tilde{z}), d_{2}(\tilde{x},\tilde{z}), f_{0}(\tilde{x}), f_{1}(\tilde{x})$ and $\lambda_{2}(\tilde{x},\tilde{z})$ have the same sign; $\gamma_{2}(\tilde{x},\tilde{z}), -\beta (\tilde{x})\eta (\tilde{x},\tilde{z})F_0(a,b) \geq 0$; and $\eta (\tilde{x},\tilde{z}), F_0(a,b) \neq 0$. \\ \hline
(Q9)& $a_{i}, b_{i} \geq 0$ for $i = 0,1,2$; $\alpha (\tilde{x}), \beta (\tilde{x}), \tilde{z}, d_{0}(\tilde{x},\tilde{z}),$ $d_{1}(\tilde{x},\tilde{z}), d_{2}(\tilde{x},\tilde{z}), f_{0}(\tilde{x}), f_{1}(\tilde{x})$ and $\lambda_{1}(\tilde{x},\tilde{z})$ have the same sign; $\gamma_{2}(\tilde{x},\tilde{z}), \beta (\tilde{x})\eta (\tilde{x},\tilde{z})F_0(a,b) \geq 0$; and $\eta (\tilde{x},\tilde{z}), F_0(a,b) \neq 0$. \\ \hline
(E1) & either (i) $\tilde{x} = 0$; (ii) $\lambda_{1}(\tilde{x}) = 0$; or (iii) $c_{0}(\tilde{x}) = 0$. \\ \hline
(E2) & either (i) $\tilde{x} = 0$; (ii) $\lambda_{3}(\tilde{x}) = 0$; or (iii) $c_{2}(\tilde{x}) = 0$. \\ \hline
(E3) & either (i) $\tilde{x} = 0$; (ii) $\lambda_{4}(\tilde{x}) = 0$; or (iii) $d_{0}(\tilde{x}) = 0$. \\ \hline
(E4) & either (i) $\tilde{x} = 0$; (ii) $\lambda_{2}(\tilde{x}) = 0$; or (iii) $d_{2}(\tilde{x}) = 0$. \\ \hline
(E5) & either (i) $\tilde{x} = 0$; (ii) $c_{0}(\tilde{x}) = 0$; or (iii) $d_{2}(\tilde{x}) = 0$. \\ \hline 
(F1) & $\lambda_{1}(\tilde{x},\tilde{z}) = 0$. \\ \hline 
(F2) & $\lambda_{3}(\tilde{x},\tilde{z}) = 0$. \\ \hline
(F3) & either (i) $\lambda_{4}(\tilde{x},\tilde{z}) = 0$, or (ii) $d_{0}(\tilde{x},\tilde{z}) = 0$. \\ \hline 
(F4) & either (i) $\lambda_{2}(\tilde{x},\tilde{z}) = 0$, or (ii) $d_{2}(\tilde{x},\tilde{z}) = 0$. \\ \hline
(F5) & $d_{2}(\tilde{x},\tilde{z}) = 0$. \\ \hline
\end{tabular}
\caption{Constraints}
\label{tab:cons}
\end{table}

\end{document}